# Granularity of algorithmically constructed publication-level classifications of research publications: Identification of specialties

Peter Sjögårde[a,b], Per Ahlgren[c,d]
[a]University Library, Karolinska Institutet, Stockholm, Sweden
[b]Department of ALM, Uppsala University, Uppsala, Sweden
[c]Department of Statistics, Uppsala University, Uppsala, Sweden
[d]KTH Library, KTH Royal Institute of Technology, Stockholm, Sweden

Email: peter.sjogarde@ki.se; per.ahlgren@uadm.uu.se

Corresponding author: Peter Sjögårde, University Library, Karolinska Institutet, 17177 Stockholm, Sweden

## Abstract

In this work, in which we build on, and use the outcome of, an earlier study on topic identification in an algorithmically constructed publication-level classification (ACPLC), we address the issue how to algorithmically obtain a classification of topics (containing articles), where the classes of the classification correspond to specialties. The methodology we propose, which is similar to the one used in the earlier study, uses journals and their articles to construct a baseline classification. The underlying assumption of our approach is that journals of a particular size and foci have a scope that correspond to specialties. By measuring the similarity between (1) the baseline classification and (2) multiple classifications obtained by topic clustering and using different values of a resolution parameter, we have identified a best-performing ACPLC. In two case studies, we could identify the subject foci of involved specialties, and the subject foci of specialties were relatively easy to distinguish. Further, the class size variation regarding the best performing ACPLC is moderate, and only a small proportion of the articles belong to very small classes. For these reasons, we conclude that the proposed methodology is suitable to determine the specialty granularity level of an ACPLC.

## 1 Introduction

In a recent article we proposed a methodology for identification of research topics in an algorithmically constructed publication-level classification of research publications (ACPLC) (Sjögårde & Ahlgren, 2018). We used a large dataset of more than 30 million publications in Web of Science to create an ACPLC, at the granularity level of topics. However, more levels of different granularity are needed for an ACPLC to be used to answer a broader range of questions. In the present study, we use a similar methodology to create a classification whose granularity corresponds to research *specialties*. In the remainder of this paper, we use the term ”specialty” instead of ”research specialty”.

The identification of specialties is part of a broader aim to develop a standard approach to create a hierarchical ACPLC of research publications in large and global, both in term of geographical uptake and coverage of subject areas, citation databases, such as Web of Science or Scopus. An ACPLC can be used for a great variety of analytical purposes and is especially useful for recurrent analytical activities.

A classification system, including a classification of publications into classes whose sizes correspond to specialties, can be used to study the publication output of different actors within a specialty, the collaboration between actors, dynamics, emergence and decline of specialties, and the relation between specialties. Moreover, a hierarchical classification, including both classes corresponding to topics and classes corresponding to specialties, makes it possible to identify topics within a specialty and, e.g., a shifting focus of a specialty. We therefore suggest that the level of specialties, together with the level of topics, should be included in a standard ACPLC, and that such an ACPLC should be hierarchical.

The purpose of this paper is to find a theoretically grounded, practically applicable and useful granularity level of an ACPLC with respect to specialties. To determine the granularity of specialties, a baseline classification is constructed. A set of journals is identified and used to create a baseline classification. ACPLCs with different granularities, constructed by the use of different values of a resolution parameter, are then compared to the baseline classification. The classification that best fits the baseline classification is proposed to be used for bibliometric analyses of specialties. In contrast to earlier work, our aim is to create a classification of publications that can be used to identify all specialties represented in Web of Science from 1980 onwards.

The remainder of this paper is structured as follows. In the next section, a short summary of our previous article on topic identification is given. The framework of the study is outlined in Section 3 and the specialty notion is discussed in Section 4. Data and methods are presented in Section 5, whereas Section 6 gives the results. Conclusions are given in Section 7.

## 2 Summary of the Sjögårde-Ahlgren study on identification of topics

To give the reader some background to the present study, we in this section summarize the earlier study on topic identification (Sjögårde & Ahlgren, 2018). In that study, we discussed how the resolution parameter given to the software Modularity Optimizer can be calibrated to obtain publication classes corresponding to the size of topics.

A set of about 31 million articles and reviews from Bibmet, KTH Royal Institute of Technology's bibliometric database, which contains Web of Science data, was used for the study. The study involved a methodology consisting of four steps. In the first step, we constructed a baseline classification ($BCP_t$) corresponding to topics, where $BCP_t$ contains synthesis articles, operationalized as articles with at least 100 references. Each such article constitutes a class, and its list of cited references points to the reference articles of the class, i.e. to the members of the class. The underlying assumption of this approach is that synthesis publications in general address a topic.

In the second step of the methodology, several ACPLCs of different granularity with respect to the topic level were created by setting the resolution parameter of Modularity Optimizer to different values. Normalized direct citation values between the articles in the dataset were used, as proposed by Waltman and van Eck (2012). For the third step, classifications derived from the ACPLCs were obtained, where each derived classification constitutes a classification of the union of the classes of the baseline classification, $BCP_t$. Thus, the latter classification and a given derived one have exactly the same underlying reference articles. In the fourth and final step of the methodology, the similarity between $BCP_t$ and each of the derived classifications from the third step was quantified. For this purpose, the Adjusted Rand Index (ARI) (Hubert & Arabie, 1985) was used. We denoted the ACPLC such that its corresponding derived classification exhibited the largest ARI similarity with $BCP_t$ by $ACPLC_t$.

With respect to the results of the study, the class size variation regarding $ACPLC_t$ turned out to be moderate, and only a small proportion of the articles belong to very small classes. Moreover, the outcomes of two case studies showed that the topics of the cases were closely associated with different classes of $ACPLC_t$, and that these classes tend to treat only one topic. We concluded that the proposed methodology is suitable to determine the topic granularity level of an ACPLC and that the ACPLC identified by this methodology is useful for bibliometric analyses.

In the present study, we use a similar methodology to identify specialties. The classes obtained in the previous study are clustered into specialties. A baseline classification is constructed that corresponds to specialties, and a set of journals is used to create the baseline classification.

We need to point out that there is a substantial overlap between our earlier paper (Sjögårde & Ahlgren, 2018) and the present one. The reason for this is that the four-step methodology used in the earlier study, and briefly described above, is used also in the study underlying the present paper.

## 3 Framework

As in the previous study, we use a network-based approach to obtain a classification of research publications (Fortunato, 2010). We use the Modularity Optimizer[1] software, created by Waltman and van Eck (2013), and the methodology put forward in Waltman & van Eck (2012). The alternative modularity function is used (Traag, van Dooren, & Nesterov, 2011), together with the SLM algorithm for modularity optimization. We acknowledge that a new algorithm for modularity optimization has been proposed (Traag, Waltman, & van Eck, 2018). However, to be consistent with the previous study, we use the SLM algorithm also in this study. We choose direct citation to express publication-publication relations, rather than bibliographic coupling (Kessler, 1965), co-citations (e.g. Marshakova-Shaikevich, 1973; Small, 1973), textual similarity (e.g. Ahlgren & Colliander, 2009; Boyack et al., 2011) or combined approaches (e.g. Colliander, 2015; Glänzel & Thijs, 2017). Direct citation is more efficient as it gives rise to fewer relations than the mentioned approaches, and there is empirical support that direct citations performs well in comparison with bibliographic coupling and co-citations when it comes to larger datasets (Boyack, 2017).

In Sjögårde & Ahlgren (2018), a network model with two levels of hierarchy, topics and specialties, was presented. This model comprises a logical classification: Each publication is classified into exactly one class at each level of hierarchy.[2] Moreover, all publications in a class, at a level below the top level, are classified into exactly one, and the same, parent class. It follows that each topic in the model belongs to exactly one specialty. In this study, in which we continue to use logical classifications, we obtain such a relation by clustering topics into specialties, rather than using the alternative approach to cluster publications directly into specialties. Logical classifications have some shortcomings: topics can be addressed by several specialties (Yan, Ding, & Jacob, 2012) or, at a higher level of aggregation, disciplines (Wen, Horlings, van der Zouwen, & van den Besselaar, 2017), phenomena not expressed by logical classifications. However, the relation between a topic and other specialties than the parent specialty, as well as relations between topics, can still be expressed and analyzed by use of the relational strengths associated with the edges in the model.

[1] http://www.ludowaltman.nl/slm/

[2] A *logical classification* of a set of objects, *O*, is a set *C* of non-empty subsets of *O* such that (a) the union of the sets in *C* is equal to *O*, and (b) the sets in *C* are pairwise disjoint. Thus, each object in *O* is classified into exactly one set in *C*.

For further discussion on the general classification framework and for an explication of a model that expresses the relations between classes at different hierarchical levels in the model, we refer the reader to Sjögårde & and Ahlgren (2018).

## 4 Specialties

Specialties have been studied since the 1960s in the field of sociology. In this literature, specialties are considered as smaller intellectual units within research disciplines (Chubin, 1976). The researchers within the same specialty communicate with each other. They possess similar competences and can engage in the same, or similar, research problems (Hagstrom, 1970). The notion of specialties is closely related to the notion of invisible colleges (Crane, 1972; Price, 1965). However, as pointed out by Morris and van der Veer Martens (2008), invisible colleges "presuppose that the researchers are in frequent informal contact with one another", which is not the case for specialties.

We use the definition of a specialty that has been given by Morris & van der Veer Martens (2008). They define a specialty as "*a self-organized network of researchers who tend to study the same research topics, attend the same conferences, read and cite each other´s research papers and publish in the same journals*". Further, and in concurrence with others, we consider specialties to be the largest homogeneous units of science "in that each specialty has its own set of problems, a core of researchers, shared knowledge, a vocabulary, and literature" (Scharnhorst, Börner, & Besselaar, 2012) and that they "play an important role in the creation and validation of new knowledge" (Colliander, 2014).

As early as 1974, Small and Griffith argued that publications can be clustered and that the obtained clusters may represent specialties (Small & Griffith, 1974). The single-linkage method was used by Small and Griffith to cluster 1,832 publications, which today would be considered a very small number of publications. They used their results to identify specialties. Since the 1970s, the technological advancements and the emergence of the Internet have changed the preconditions for research communication. There has also been a growth in research activity and production of research publications.

More lately, specialties have been identified and analyzed by the use of different clustering techniques (Lucio-Arias & Leydesdorff, 2009; Morris & van der Veer Martens, 2008; Scharnhorst et al., 2012). Different points of departure and different operationalizations of the specialty notion have captured different aspects of specialties. For example, clustering of publications based on citation relations and clustering of researchers based on co-authorship may result in different pictures of a specialty. The former approach identifies a set of publications and the latter a group of researchers belonging to a specialty. We attempt to capture the *publications* belonging to each specialty, rather than the *researchers* belonging to the specialty. A researcher can be part of several specialties, a property that cannot be expressed by the co-authorship approach. For this reason, we consider this approach less suitable for the identification of publications belonging to a specialty. We believe that it is preferable to base classifications constructed for the purpose of bibliometric analyses of specialties on the network of publications, rather than on the network of researchers. Our approach makes it possible to identify the researchers within a specialty without forcing every researcher into exactly one specialty. It also makes it possible to analyze the contribution of one researcher to multiple specialties.

Kuhn (1996) estimates the number of core researchers in a specialty to be around 100. Based on Lotka's law (1926), Morris (2005) estimates the total size of researchers within a specialty to be around 1,000, and the number of publications produced by a specialty to be between 100 and 5,000. Boyack et al. (2014) regard specialties to be "ranging from roughly a hundred to a thousand articles per year." We acknowledge that the size of specialties in terms of publications may vary over time. Because the output of research publications have been growing the last decades, it is likely that the total size of specialties, in terms of number of publications, has been growing. Also the yearly publication production of active specialties are likely to be on average larger today than ten or twenty years ago. The size of specialties is an empirical question that we intend to shed light on in the present study.

# 5 Data and methods

As in Sjögårde and Ahlgren (2018), KTH Royal Institute of Technology's bibliometric database Bibmet was used for the study. Bibmet contains Web of Science publications from the publication year 1980 onwards. In the present study, we use the same set of publications as in the earlier study. We denote this set, in agreement with the earlier study, by $P$. $P$ consists of 30,669,365 publications of the two document types Article" and "Review". In the remainder of this paper, we use the term "article" to refer to both articles and reviews.

## 5.1. Design of the study

We attempt to find a granularity of an ACPLC, where the ACPLC is based on the articles in $P$, that correspond to specialties. In order to identify the granularity of specialties, a baseline classification of publications (BCP) is created. The BCP is a set of journals, considered as classes, and each member of a class in BCP is a publication appearing in the class, i.e. appearing in the journal.

The BCP is compared to several ACPLCs with different granularities, where each such ACPLC is obtained by clustering the classes of $ACPLC_t$ (see Section 2), which is thereby utilized in the present study. An appropriate granularity is detected and an ACPLC is chosen, the classes of which correspond to specialties. The methodology, which has four steps and a high degree of similarity with the methodology proposed in Sjögårde and Ahlgren (2018), is described in detail in step I to IV below and schematically illustrated in Figure 1.

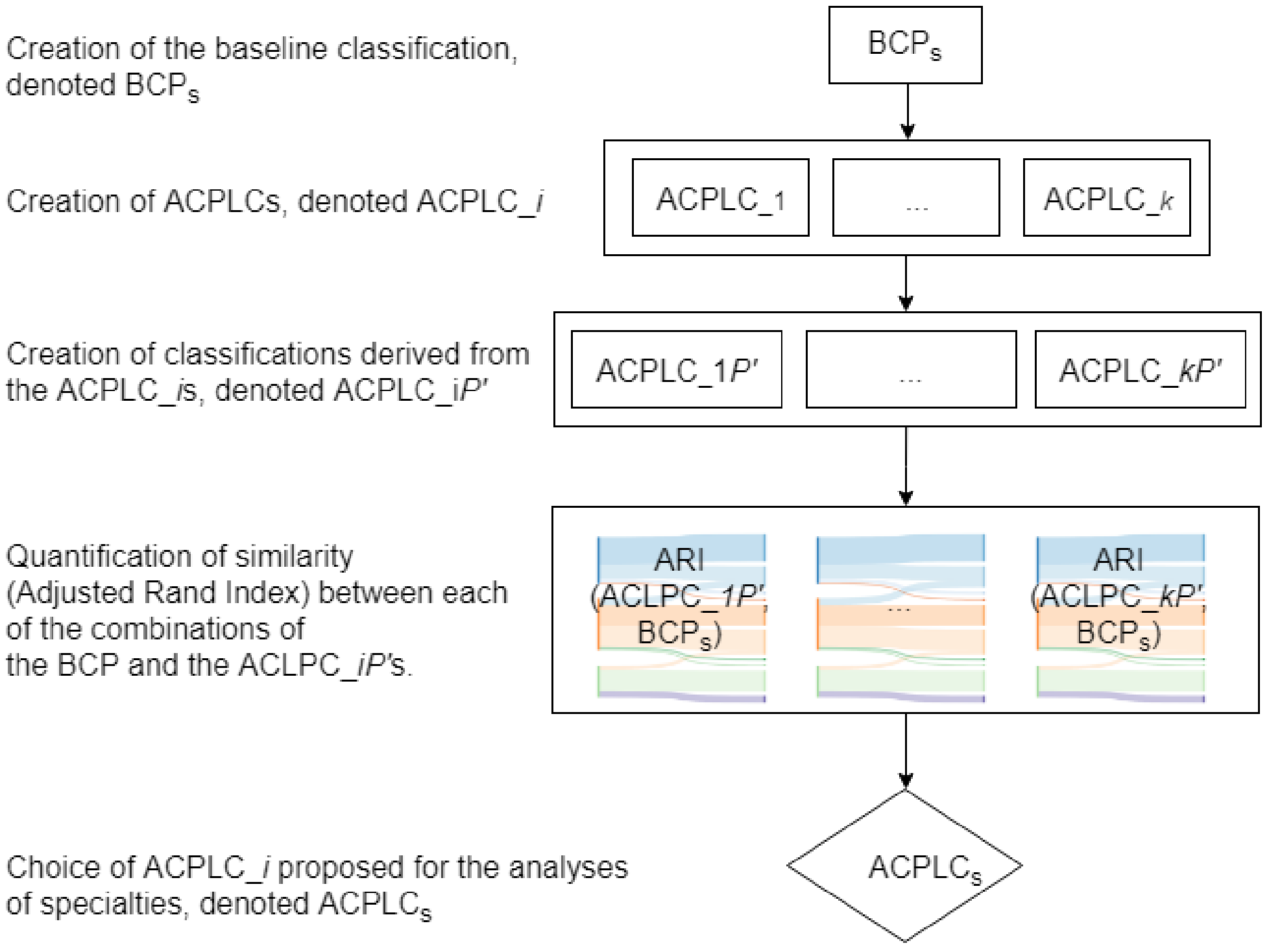

Figure 1: Illustration of the design of the study.

## *I. Creation of baseline classes*

We construct a baseline classification to correspond to specialties, which we denote by $BCP_s$. For the creation of $BCP_s$, a subset of journals covered by Web of Science is used. Each journal constitutes a class, and the publications appearing in the journal are the members of the class.

The reason to use journals to obtain $BCP_s$ is that researchers within a specialty publish in and read the same journals. The new possibilities to search, retrieve and read research articles have changed the role of journals, nevertheless many journals are still focused on specific areas of expertise and the researchers within those areas. Such journals aim to publish articles that are relevant to its audience. E.g. we consider bibliometrics as a specialty within the discipline of library and information science, and the scope of *Journal of Informetrics* as roughly targeting the specialty of bibliometrics. In resemblance with Bradford's law (1948), a researcher within a specialty needs to go to several journals to find all relevant articles within her or his specialty. The boundaries of a specialty are vague and fading rather than sharp. If we consider a journal, which scope roughly covers a specialty, a core set of the articles in such journal is likely to be of high relevance to the core audience of the journal. The researchers that belong to this core audience can be considered as the backbone of the specialty. The rest of the articles in the journal have a fading relevance to this specialty. Some of these articles will be of higher relevance to other specialties.

When creating $BCP_s$, we attempt to delimit the set of journals to such journals that, regarding their size and scope, can be considered as proxies for specialties. Since $BCP_s$ is to be used as a baseline to estimate granularity of an ACPLC regarding specialties, the following three requirements should be addressed:

A. To be able to compare the classifications, the union of the classes in $BCP_s$ must be a subset of the union of the classes (i.e. the topics) in $ACPLC_t$.
B. Ideally, each class (journal) in $BCP_s$ should address exactly one specialty.
C. Ideally, each pair of distinct classes (journals) should address different specialties.

Now, as a first step to satisfy point A, we kept, for a given journal, only articles, i.e. publications that are of the document types "Article" or "Review". We return to this point in the end of this subsection.

To deal with point B, we first delimited the publication period to one year, namely 2010. By this operation, which resulted in 12,276 journals, the risk of including journals that, for instance, have shifted subject focus over time is lowered. In addition to deal with point B, the choice of publications from a publication year that have both incoming and outgoing citations can be assumed to have a stabilizing effect when these articles are being clustered, compared to more recent publications.

We then removed all journals belonging to the Web of Science subject category "Multidisciplinary Sciences", since a journal in this category is clearly not focused to a single specialty. After this, 12,233 journals remained. Next, we considered the distribution of articles over journal size. Figure 2 shows the distribution limited to journals with less than or equal to 1,000 articles. A typical article is published in a journal that, with respect to size and modal interval as a measure of central tendency, published 30-40 articles in 2010. By including journals between the 5th and 50th percentiles of the journal size distribution displayed in Figure 2, journals with 28-194 articles were included. With this journal size limitation, the risk to include journals addressing multiple specialties (or journals with a narrower scope than a specialty) is reduced. The limitation reduced the number of journals to 7,481.

Finally, in order to further reduce the risk of including journals addressing multiple specialties, we took journal self-citations into account. The idea is that a one-specialty journal can be assumed to cite itself to a larger extent compared to a journal that covers two or more specialties, other things held constant. In the light of this, we required, for a journal to be included in $BCP_s$, that the self-citation ratio (in %) should be at least 10.[3] This further reduced the number of journals to 1,404.

Some of the measures taken to satisfy point B are also relevant for satisfying point C (which states that each pair of distinct classes should address different specialties), for instance the limitation to the publication year 2010. With the aim to further raise the possibilities to satisfy point C, we applied bibliographic coupling between journals. If two journals had an overlap of 8% or more regarding their active cited references, they were considered as specialty overlapping.[4] The threshold was chosen after browsing a list of journal pairs ordered descending after number of shared cited references. We grouped journals so that all journals that were directly or indirectly connected, by a cited reference overlap of 8% or more, were assigned the same group. E.g. if journal *j*1 has an cited reference overlap of ≥ 8% with journal *j*2, and *j*2 has an cited reference overlap of ≥ 8% with *j*3, then *j*1, *j*2 and *j*3 are assigned to the same group. Note that *j*1 and *j*3 are assigned to the same group, even if they do not have an active reference article overlap of ≥ 8%. Each obtained group of journals was considered as addressing the same specialty. One of the journals was then randomly selected from each group. After the execution of this procedure, 1,119 journals remained. This number is the number of journals (classes) in $BCP_s$.

The number of articles belonging to the classes of $BCP_s$ was initially 84,139. However, only articles that belong to a class in $ACPLC_t$, the best-performing ACPLC in the topic study, were kept, in order to satisfy point A above. Thereby, 7% of the articles were removed, and therefore the number of articles in $BCP_s$ is 78,217. Note that articles not present in $ACPLC_t$ lack citation relations. We denote the union of the classes in $BCP_s$ as *P'*.

---

[3] The *self-citation ratio* (*s*) for a journal *j* is given by:

$$s_j = \frac{c_s}{r_a} \quad (1)$$

where $c_s$ is the number of self-citations in *j*, and $r_a$ the total number of active references in *j*. References are considered as active if they point to publications covered by the data source (Waltman, van Eck, van Leeuwen, & Visser, 2013). A reference is considered as a self-citation if the referencing publication and the referenced publication belong to the same journal.

[4] The *overlap* (*y*) between two journals ($j_1$ and $j_2$) is given by:

$$y = \frac{1}{2}\left(\frac{m}{A_1} + \frac{m}{A_2}\right) \quad (2)$$

where *m* is the number of shared cited references, i.e. cited references occurring in both $j_1$ and $j_2$, $A_1$ the number of cited references in $j_1$ and $A_2$ the number of cited references in $j_2$. The reference list of a journal was obtained by concatenating the reference lists of the articles (published year 2010) in the journal. If a reference article has been cited by more than one article in a journal, then this reference is counted multiple times for that journal. E.g. if journal *j*1 has four references to article *a* and journal *j*2 has two references to article *a*, than journal *j*1 and *j*2 have two shared cited references with respect to article *a*. Note that we give the overlap measure threshold as a percentage in the running text.

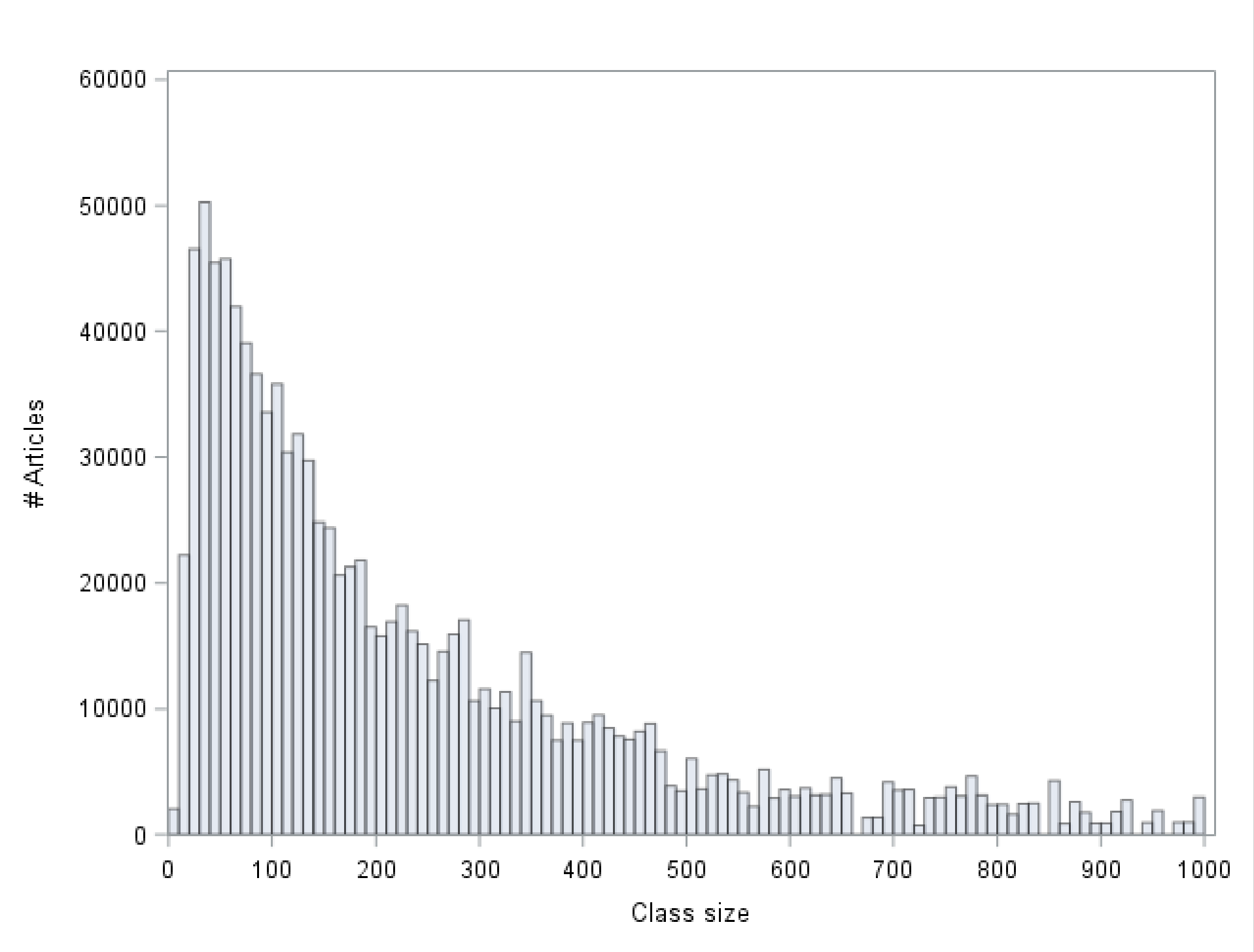

Figure 2: Number of articles per journal size for journals with 1 to 1,000 articles in 2010.

*II. Creation of ACPLCs of different granularity with respect to the specialty level*

In order to obtain ACPLCs of different granularity, the first step was to measure the relatedness between the classes (topics) of $ACPLC_t$. We measured the relatedness as the average normalized direct citation value between the articles belonging to the two classes: If class $C$ contains $m$ articles and class $C'$ $n$, the sum of the $m \times n$ normalized direct citation values between articles in $C$ and articles in $C'$ was divided by $m \times n$. In the second step, the generated class relatedness values were iteratively given as input to Modularity Optimizer to cluster the classes of $ACPLC_t$, where the resolution parameter was set to different values in the iterations. By this, ACPLCs were created for comparison of similarity with $BCP_s$. We denote the ACPLCs by ACPLC_1, …, ACPLC_$k$, where $k$ is the number of created ACPLCs.

*III. Creation of classifications derived from the ACPLCs*

For each ACPLC_$i$ ($1 \leq i \leq k$), a classification was derived from ACPLC_$i$ in the following way:

(a) Each class $C$ in ACPLC_$i$ such that $C$ did not contain any articles in $P$' was removed from ACPLC_$i$. Let ACPLC_$i$1 be the subset of ACPLC_$i$ that resulted from the removal.
(b) For each class $C$ in ACPLC_$i$1, all articles in $C$ that did not belong to $P$' were removed from $C$. Let ACPLC_$iP'$ be the set that resulted from these removal operations.

Clearly, the set ACPLC_$iP'$ constitutes a classification of $P$', i.e. of the union of the classes of the baseline classification $BCP_s$. Thus, ACPLC_$iP'$ and $BCP_s$ have exactly the same underlying articles. We denote the $k$ derived classifications as ACPLC_1$P'$, …, ACPLC_$kP'$. These classifications then correspond to the classifications ACPLC_1, …, ACPLC_$k$.

*IV. Quantification of the similarity between $BCP_s$ and the ACPLC_iP's*

We attempt to optimize the granularity of an ACPLC_$iP'$ so that it exhibits as high similarity as possible with $BCP_s$. Figure 3 illustrates the relation between two classifications as an alluvial diagram. Example $A$ shows two classifications $A_1$ and $A_2$ with a high similarity. Example $B$ shows two classifications where one of the classifications is more coarsely grained ($B_1$) than the other classification ($B_2$). The similarity between $A_1$ and $A_2$ is higher than the similarity between $B_1$ and $B_2$. If we consider $B_1$ as a baseline classification, then the granularity of $B_2$ would be too finely grained.

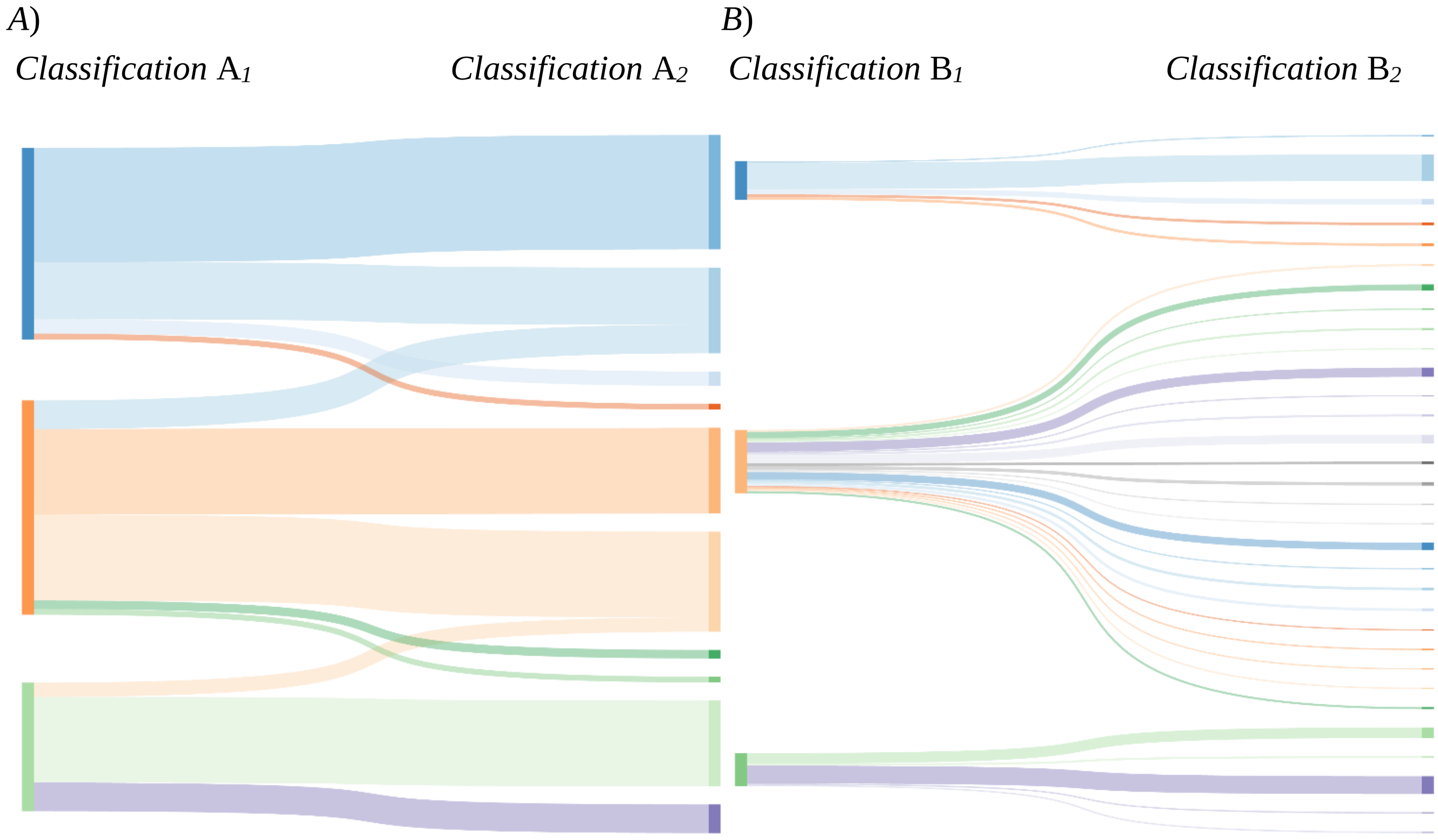

Figure 3: Two alluvial diagrams (A and B) illustrating the relation between two classifications. A shows two classifications with a high level of similarity. B shows two classifications with a low level of similarity.

As in our topic identification study, we used the Adjusted Rand Index (ARI) (Hubert & Arabie, 1985) to quantify the similarity between BCP$_s$ and an ACPLC_*iP’*. The ARI ranges from 0 to 1. It is advantageous over the original Rand Index proposed by Rand (1971), because it adjusts for chance. The ARI compares two classifications by considering pairs of items in one of the classifications and whether or not each pair is grouped into the same class in the other classification. Note that an ARI value of 1 between BCP$_t$ and an ACPLC_*iP’* corresponds to a situation in which these two classifications are identical. For further information on ARI, we refer the reader to Sjögårde and Ahlgren (2018).

To find the ACPLC_*iP’* with the highest ARI similarity with BCP$_s$, we tested the similarity after each run of Modularity optimizer. A first run was made with a resolution parameter value of 5E-7. This value was chosen based on previous experience and some testing. We then increased the parameter value with 5E-7. This increase resulted in a higher ARI similarity, and we therefore increased the resolution further with 5E-7 for the third run, from 1E-6 to 1.5E-6. We continued by increasing the resolution by 5E-7, in total three more times, and thus six runs were done. The fourth run, with a resolution parameter value of 2E-6, gave rise to the highest ARI similarity (Table 2 and Figure 4, Section 6).

In total BCP$_s$ consists of 1,118 baseline classes. A given ACLPC_*iP’* consists of 78,217 articles, which is about 7% of the articles from the year 2010 in the corresponding ACPLC_*i*. The ACPLC_*i* such that ACLPC_*iP’* exhibits the largest ARI similarity with BCP$_s$ is proposed to be used for the analyses of specialties. We denote this ACPLC_*i* by ACPLC$_s$.

# 6 Results and discussion

In this section, we first deal with the selection and properties of $ACPLC_S$. Then, as in the earlier study on topic identification (Sjögårde & Ahlgren, 2018), we consider two cases. We examine the specialties of articles belonging to (1) the Web of Science subject category "Information science & Library Science", and (2) the Web of Science subject category "Medical Informatics".

## 6.1. Selection and properties of $ACPLC_S$

Figure 4 shows a scatter plot of the relation between the resolution value (horizontal axis) used to obtain ACPLC_*i*s and the ARI value (vertical axis), obtained by comparing the ACPLC_*iP'*s with $BCP_S$. ACPLC_4*P'* has the highest ARI value. ACPLC_4*P'* corresponds to ACPLC_4, which we consider to be the most proper ACPLC_*i* with respect to granularity of specialties. In the remainder of this paper, we denote ACPLC_4 as $ACPLC_S$. However, we acknowledge that ACPLC_3*P'* has an ARI value that is only slightly lower than the value of ACPLC_4*P'*. Thus, ACPLC_3*P'* performs almost as good as ACPLC_4*P'*.

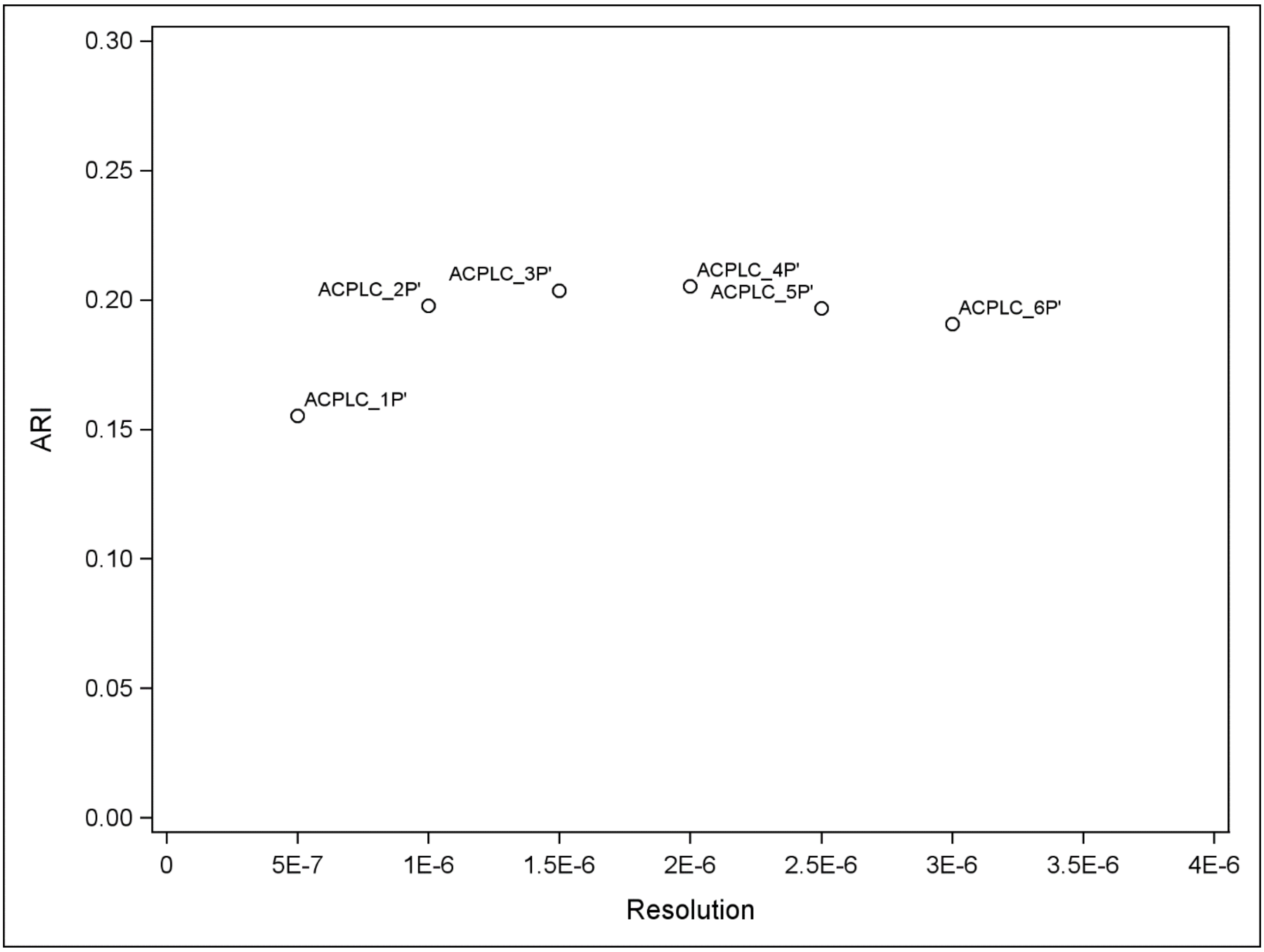

Figure 4: ARI values between ACPLC_iP's and $BCP_S$. The vertical axis shows the ARI value and the horizontal axis shows the value of the resolution parameter used to obtain the corresponding ACPLC_*i*s. The order of ACPLC_i*P'*s corresponds to their order in Table 2.

To get a picture of how well $ACPLC_S$ matches $BCP_S$, we calculated the distribution of articles in an average class in $BCP_S$ into classes (journals) in $ACPLC_S$. This was done by first calculating the average number of classes in $ACPLC_S$ into which the articles in a class in $BCP_S$ are distributed, an average that is equal to 24 (after rounding to nearest integer). We then selected all 19 classes in $BCP_s$ that were distributed into exactly 24 classes. Let the set of these classes be $P_{sc}$. The average number of articles in a $P_{sc}$ class is 63. For each of the $P_{sc}$ classes, we calculated the number of its articles in each of the 24 $ACPLC_s$ classes and sorted the resulting table in descending order. The $ACPLC_s$ class with the highest number of articles (i.e. the class corresponding to the first row in the table) was assigned the rank 1, the second largest class (i.e. the class corresponding to the second row in the table) was assigned the rank 2, etc. In this way, 19 ranked tables were obtained. Finally, averages of the number of articles by rank number, 1,…, 24, were calculated across all the 19 tables. Figure 5 shows the resulting average distribution of articles in $P_{sc}$ (to the left) into the 24 $ACPLC_t$ classes (to the right). Ranks and average number of articles across the $P_{sc}$ classes are shown for $ACPLC_s$.

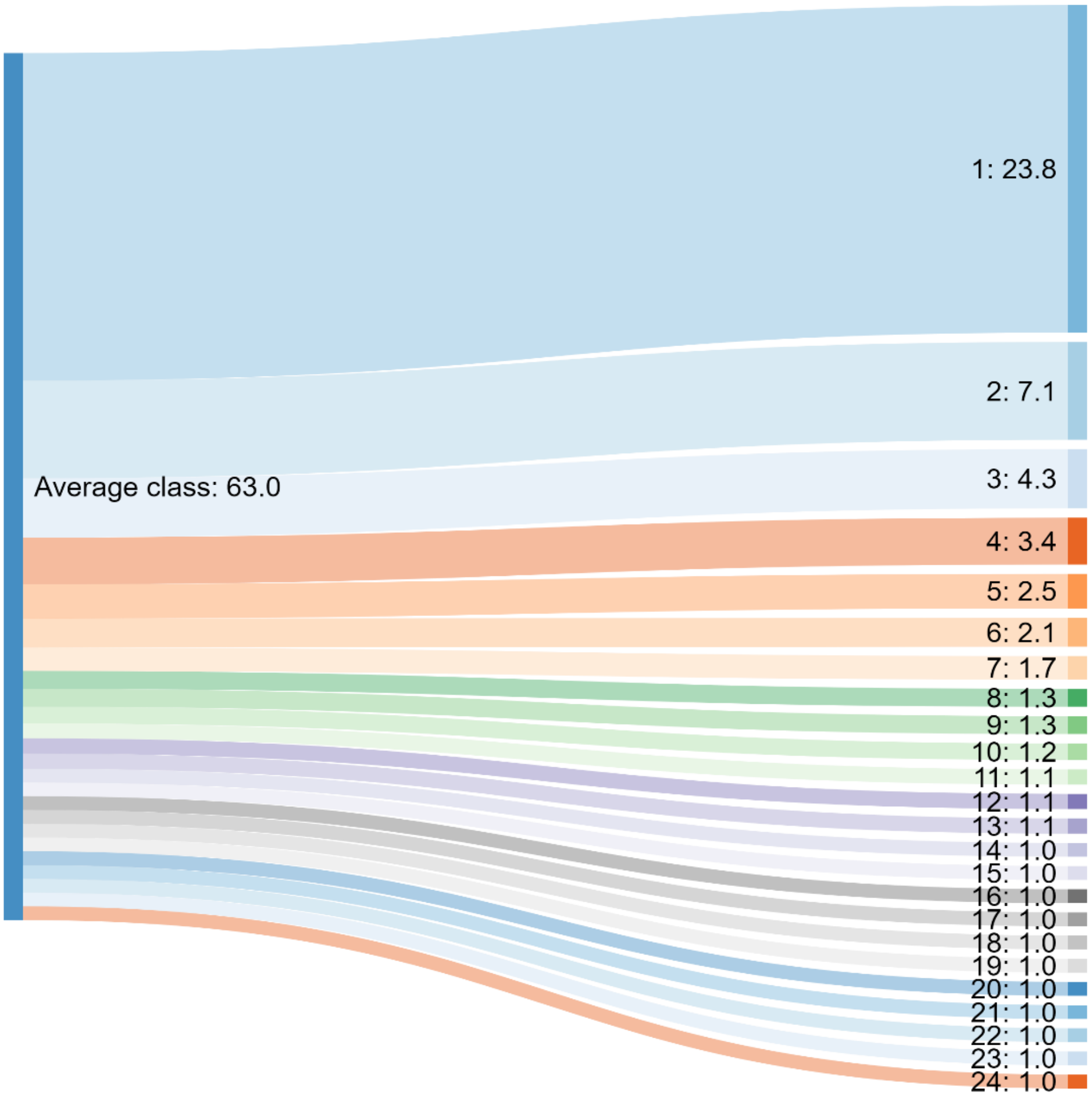

Figure 5: Alluvial diagram for an average class. The diagram shows the distribution of journal articles in $BCP_s$ into $ACPLC_s$.[5]

Given that we consider the classes in $ACPLC_s$ as specialties, the distribution of journal articles in a typical $BCP_s$ class follows a skewed distribution of specialties. About 49% of the articles in an average $BCP_s$ class are distributed into the two most frequent specialties, and 17 specialties (classes 8 to 24) are represented by a single article (after rounding to nearest integer). Hence, a high share of the articles of the average $BCP_s$ class is concentrated to a few of the $ACPLC_s$ classes. We therefore consider the match between $ACPLC_s$ and $BCP_s$ as good.

[5] http://sankeymatic.com/ has been used for the illustration.

ACPLC$_s$ consists of 60,649 classes, ranging from 1 to 60,608 articles. Most of the classes are small in size, however these classes contain a small share of the total number of articles in ACPLC$_s$. For instance, classes with less than 500 articles contain about 0.9% of the publications in ACPLC$_s$. Figure 6 shows a histogram of the distribution of classes by class size (in terms of number of articles). In order to include classes of a substantial size in the figure, classes with less than 500 articles has been excluded in the figure.

Most specialties of substantial size (minimum of 500 articles) have 7 (10$^{th}$ percentile) to 90 (90$^{th}$ percentile) subordinated topics of substantial size (a minimum of 50 articles), with a mode value of 10, a median of 28 and a mean value of about 40 (Figure 7 and Table 1).

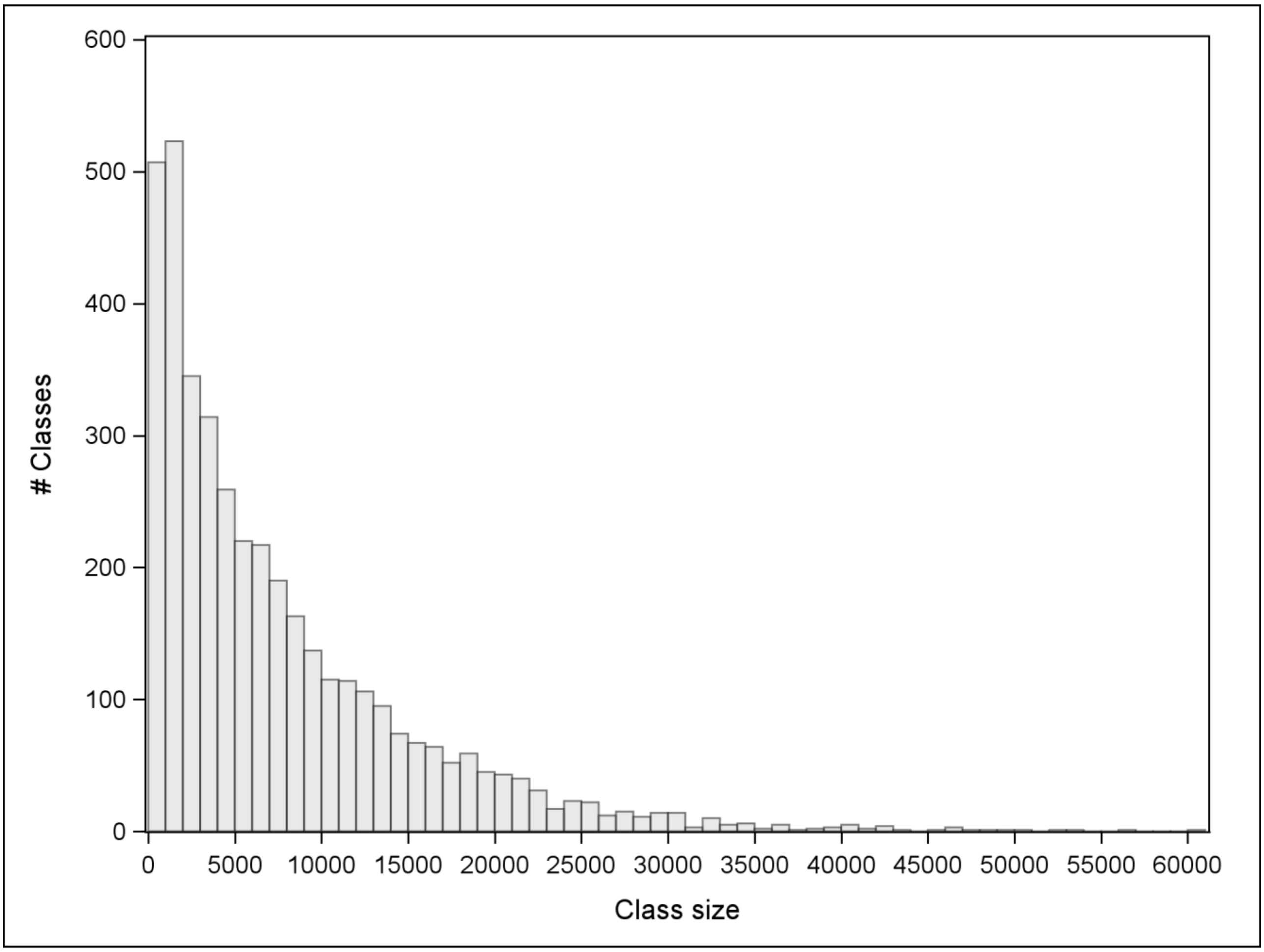

Figure 6: Histogram of number of classes by class size for ACPLC$_s$. Classes with less than 500 articles disregarded.

Table 1: Distribution statistics of number of topics per specialty for ACPLC$_s$. Specialties with less than 500 articles and topics with less than 50 articles disregarded.

| **Mean # topics per specialty** | **Median # topics per specialty** | **Mode # topics per specialty** | $P_{10}$ | $P_{90}$ |
|---|---|---|---|---|
| 39.8 | 28 | 10 | 7 | 90 |

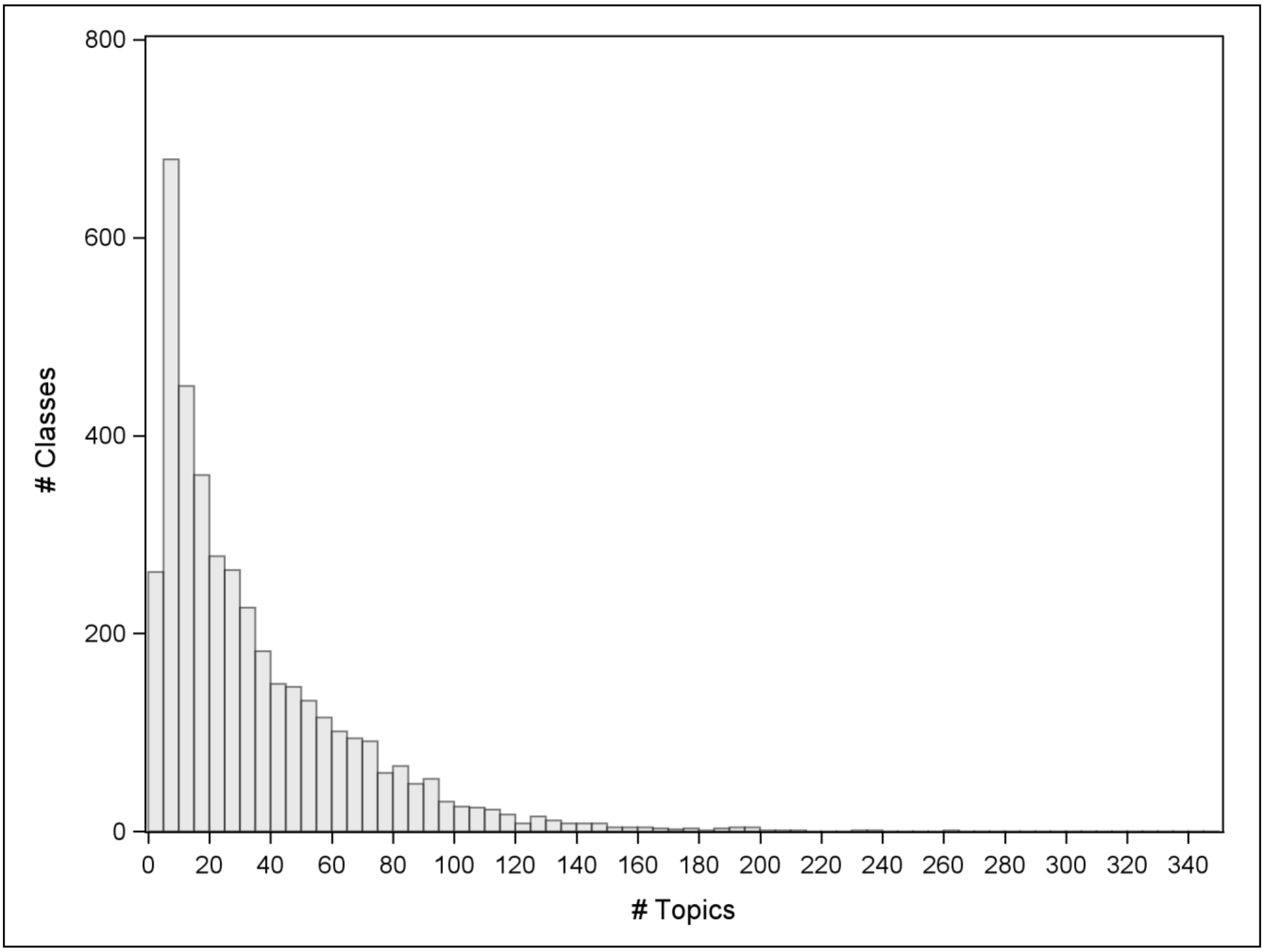

Figure 7: Histogram of number of specialties by number of subordinated topics for $ACPLC_S$. Specialties with less than 500 articles and topics with less than 50 articles disregarded.

In Figure 8 class sizes are plotted by rank order for $ACPLC_S$ (= ACPLC_4), as well as for ACPLC_3 and ACPLC_5. A log-10 scale is used on both the vertical axis (showing class size by number of articles) and the horizontal axis (showing ranks). In this figure, all classes are shown, including small size classes. About 3,500 classes contain at least 1,000 articles, about 1,100 classes contain at least 10,000 articles and about 70 classes contain at least 30,000 articles. In agreement with our study on topics, the size of classes is dropping rather slowly, regardless of classification. The increasing granularity–from ACPLC_3 via $ACPLC_S$ to ACPLC_5–is reflected by, for example, corresponding, decreasing intercepts.

Figure 9 expresses the number of articles in *P* (vertical axis) that are associated with different class sizes (horizontal axis). For a randomly selected article *a*, it is most probable that the size of the specialty class in $ACPLC_S$ to which *a* belongs is 7,000-8,000 articles (cf. the highest bar of the histogram in Figure 9). 80% of the articles belong to classes consisting of 3,765 ($10^{th}$ percentile) to 29,509 ($90^{th}$ percentile) articles (Table 2). The median value of $ACPLC_S$ is 13,145 and the mean 15,228. This distribution is not as skewed as the corresponding topic distribution (Sjögårde & Ahlgren, 2018, Figure 8).

The number of articles contributing to a specialty in 2015 (the most recent complete year at the time for data extraction) is between 187 and 2,040, given that we only take the mid 80% of the distribution into account (Table 3 and Figure 10). The median class size is 742. The mean number of articles per specialty class is growing approximately linearly across the 10 years (Table 3). This can be expected, considering the linear growth of research publication output in Web of Science.

As mentioned in the introduction, Morris (2005) estimates the size of specialties to be between 100 and 5,000 articles, however not mentioning any time period, and Boyack et al. (2014) estimate the yearly article output of a specialty to be somewhere between 100 and 1,000 articles. The results of the present study cannot be easily compared to these figures. Both the estimation of Morris and Boyack et al. are rough. Morris does not mention any time period. Further, the work by Morris is rather old and the size of specialties may have increased, in terms of publication output. Table 3 shows that the number of articles in Web of Science has been growing by more than 50% between 2006 and 2015. In 2015, the size of specialties range from about 200 articles (10$^{th}$ percentile) to 2,000 (90$^{th}$ percentile) articles. Thus, the size of specialties in 2015 is about twice the size estimated by Boyack et al. We regard this difference as rather small, taking into account that Boyack et al. defines the next larger level (disciplines) to range from tens to hundreds of thousand articles per year, thus several orders of magnitude larger than our estimation of the size of specialties.

In agreement with Morris and Boyack et al., we find it reasonable not to consider publication classes under some threshold to be regarded as specialties. One solution to the problem of small class sizes is to reassign such classes (classes below a threshold) based on their relations with larger classes (classes above or equal to the same threshold) as proposed by Waltman & van Eck (2012). However, how to set the threshold is a question that we do not address in this paper.

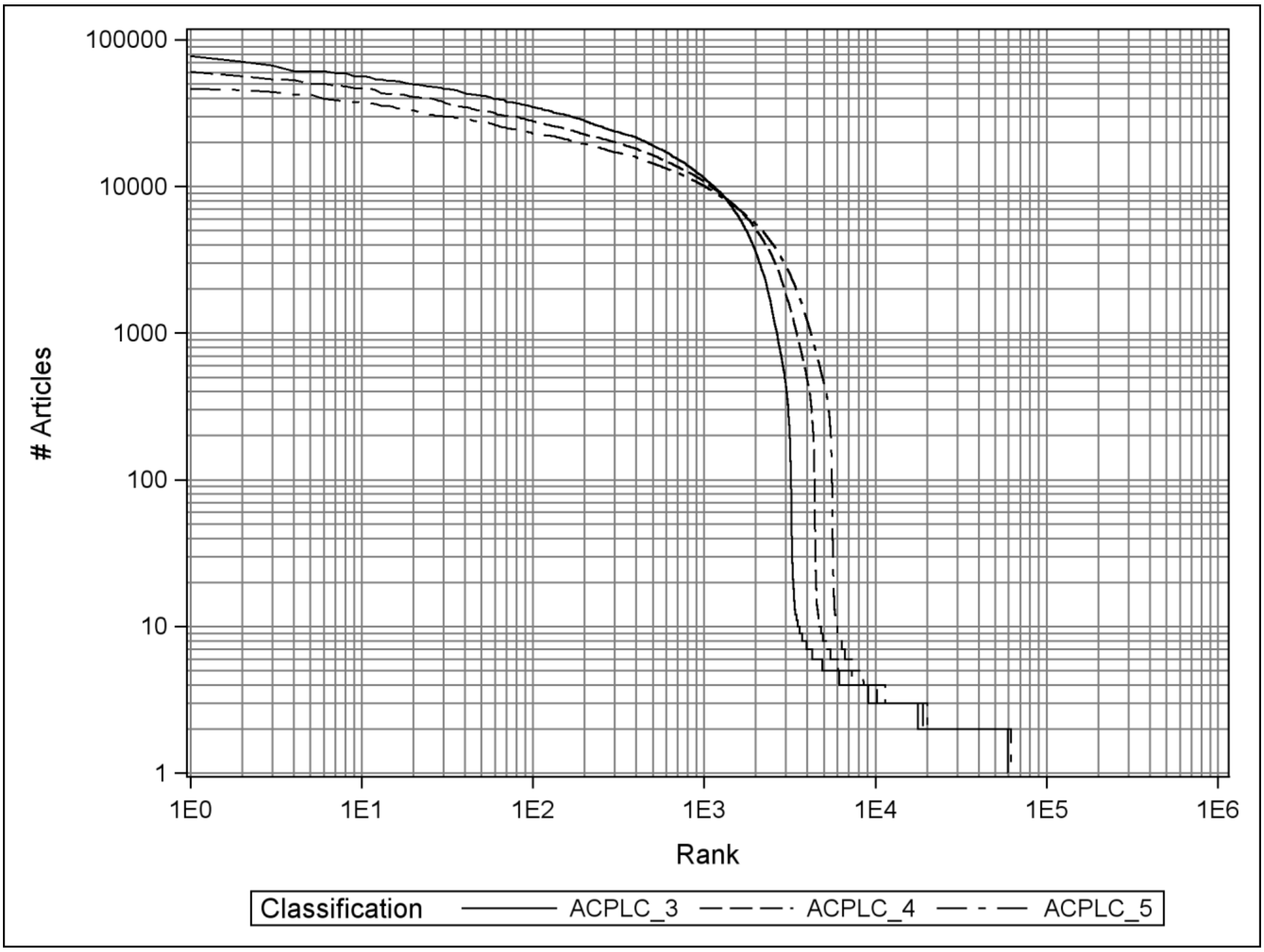

Figure 8: Distribution of number of articles by class size for three classifications. The classes in ACPLC_3, ACPLC_4 = $ACPLC_s$ and ACPLC_5 are ordered descending by size with respect to the horizontal axis. Log-10 scale used for both axes.

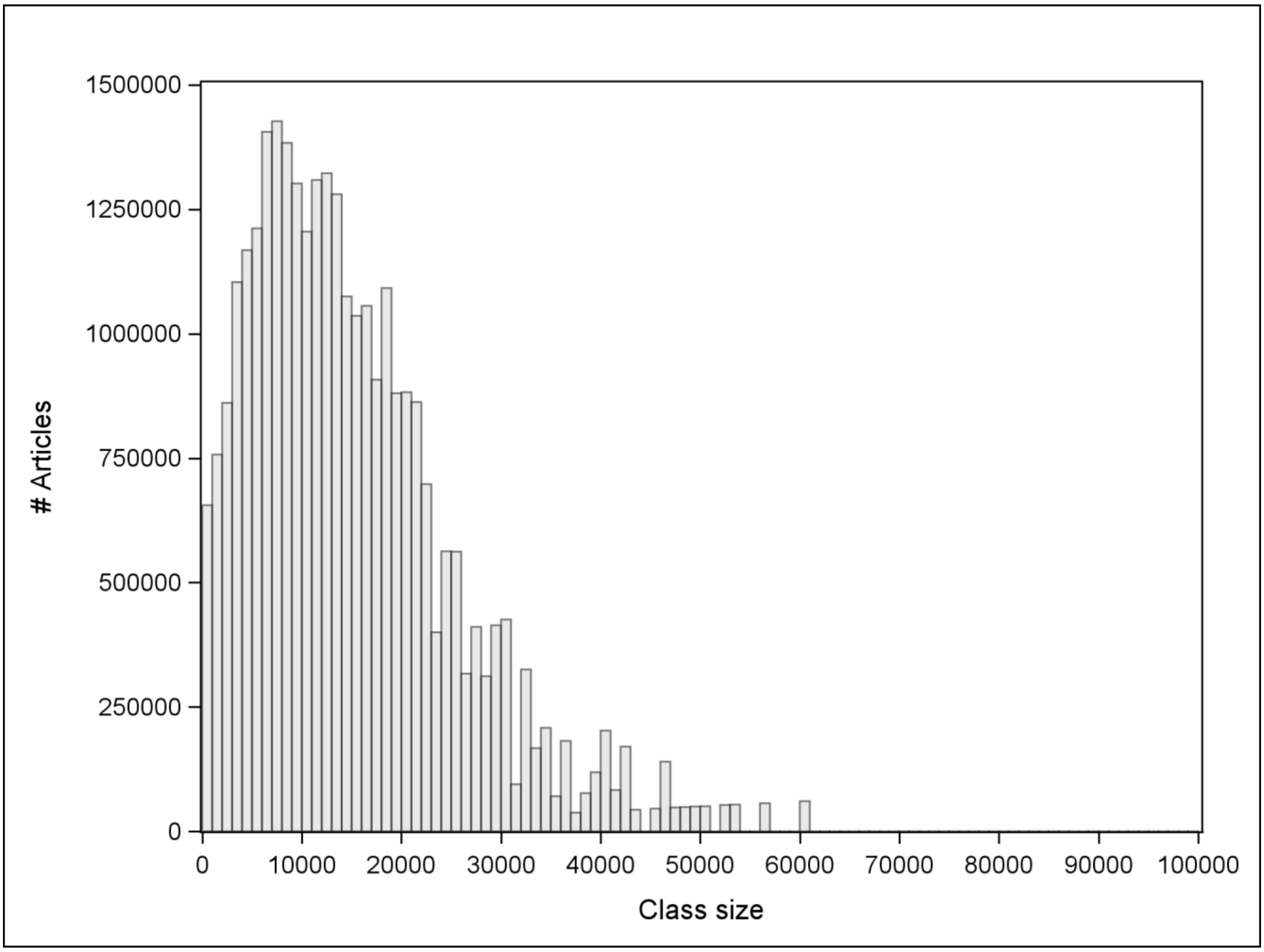

Figure 9: Histogram of number of articles by class size for $ACPLC_S$.

Table 2: For each ACPLC_iP’, the ARI value between ACPLC_iP’ and $BCP_S$, and the value of the resolution parameter used to obtain ACPLC_i, are shown, as well as number of classes with at least 500 articles and class size distribution measures for ACPLC_i.

| | | | | **Weighted class size distribution measures regarding ACPLC_*i* (*i* = 1, …, 6): Mean, Median, 10th and 90th percentile (denoted $P_{10}$ and $P_{90}$)** | | | |
|---|---|---|---|---|---|---|---|
| **Denotation** | **Resolution** | **ARI value** | **# classes with # articles ≥ 500** | **Mean # articles per class** | **Median # articles per class** | $P_{10}$ | $P_{90}$ |
| ACPLC_1*P’* | 0.0000005 | 0.155 | 881 | 66,750 | 57,984 | 19,552 | 121,981 |
| ACPLC_2*P’* | 0.0000010 | 0.198 | 1,888 | 31,123 | 27,377 | 8,866 | 59,985 |
| ACPLC_3*P’* | 0.0000015 | 0.204 | 2,953 | 20,426 | 17,960 | 5,260 | 39,326 |
| ACPLC_4*P’* | 0.0000020 | 0.205 | 3,969 | 15,228 | 13,145 | 3,765 | 29,509 |
| ACPLC_5*P’* | 0.0000025 | 0.197 | 4,897 | 12,016 | 10,499 | 2,899 | 22,819 |
| ACPLC_6*P’* | 0.0000030 | 0.191 | 5,770 | 9,936 | 8,589 | 2,342 | 18,655 |

Table 3: For a 10 year period (at the time for data extraction), the table shows class size distribution measures for $ACPLC_S$.

| | | Weighted distribution measures regarding $ACPLC_S$: Mean, Median, 10th and 90th percentile (denoted $P_{10}$ and $P_{90}$) | | | |
|---|---|---|---|---|---|
| **Publication year** | **# Articles** | **Mean # articles per class** | **Median # articles per class** | $P_{10}$ | $P_{90}$ |
| 2006 | 989,420 | 555 | 465 | 126 | 1,082 |
| 2007 | 1,040,026 | 584 | 480 | 132 | 1,150 |
| 2008 | 1,115,118 | 630 | 513 | 141 | 1,234 |
| 2009 | 1,166,665 | 667 | 541 | 145 | 1,351 |
| 2010 | 1,210,495 | 704 | 568 | 150 | 1,424 |
| 2011 | 1,290,309 | 764 | 613 | 159 | 1,545 |
| 2012 | 1,358,175 | 819 | 652 | 168 | 1,653 |
| 2013 | 1,435,835 | 890 | 697 | 177 | 1,778 |
| 2014 | 1,478,273 | 945 | 723 | 186 | 1,901 |
| 2015 | 1,524,010 | 995 | 742 | 187 | 2,040 |

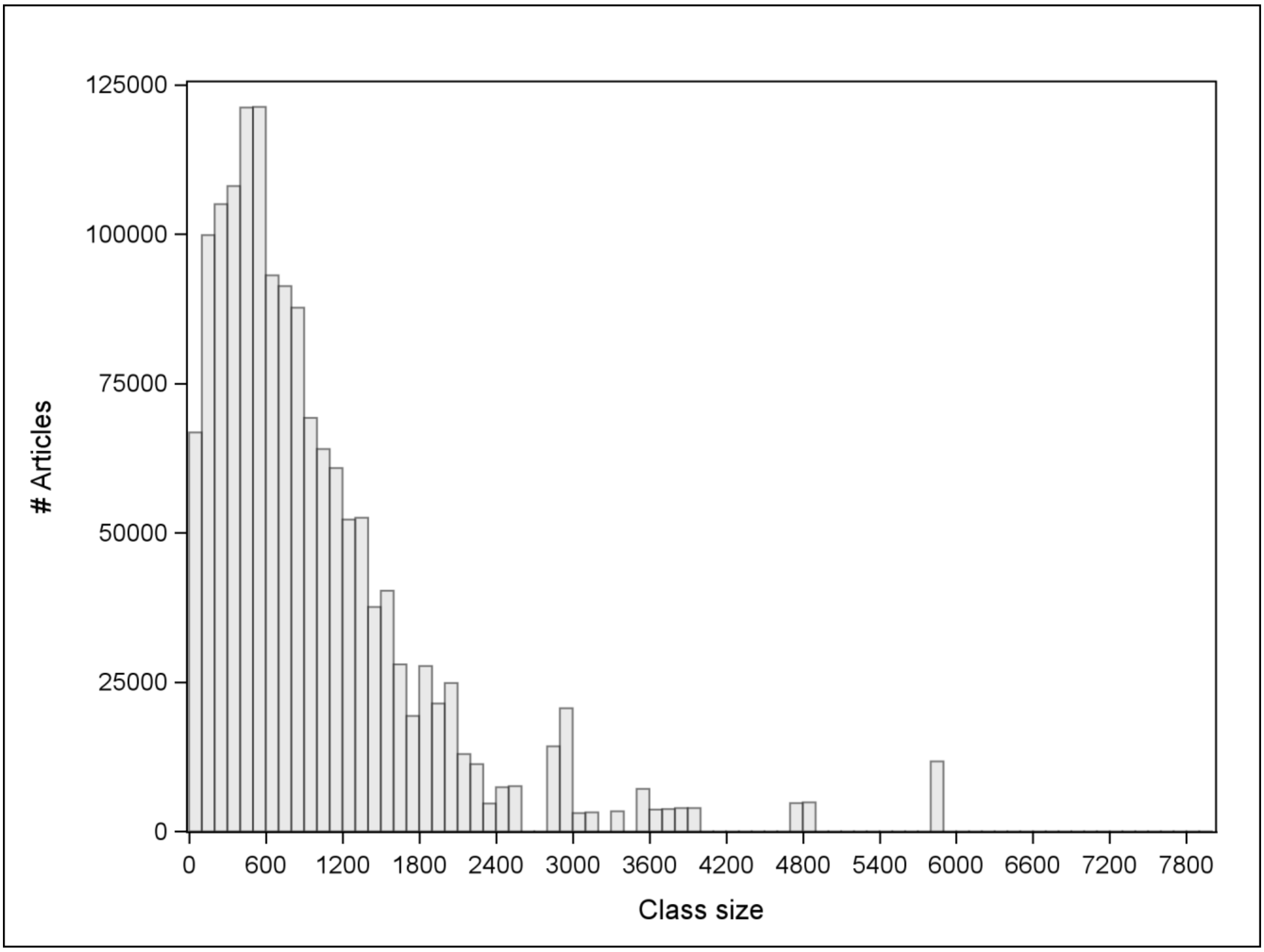

Figure 10: Histogram of number of articles by class size, for the publication year 2015 and for $ACPLC_S$.

## 6.2. The case of Library and Information Science

To explore how articles within the discipline of library and information science (LIS) are distributed into classes in $ACPLC_s$, we retrieved all articles in *P* that belong to a journal classified into the Web of Science subject category "Information science & Library Science" and published in the period 2011-2015. In total, 16,278 articles were retrieved. Let $P_{lis}$ be this set of articles.

For each class in $ACPLC_s$, labels were automatically created based on author keywords (Sjögårde & Ahlgren, 2018). To distinguish the scope of each specialty, we used these labels and the labels of the topics in each class. Recall that $ACPLC_s$ is obtained by clustering the topics of $ACPLC_t$, the best performing ACPLC with respect to topic identification (Sjögårde & Ahlgren, 2018). Table 4 shows the total number of articles in the 10 most frequent specialties and the number, and the share, of articles in a specialty that belong to $P_{lis}$. The top 10 specialties cover about 55% of the articles in $P_{lis}$. Some of the top ten specialties are highly concentrated to the analyzed Web of Science subject category (e.g. "ACADEMIC LIBRARIES//INTERLENDING//DOCUMENT DELIVERY", 77%), while other specialties have a low share of its total number of articles in this category (e.g. "CUSTOMER SATISFACTION//CUSTOMER LOYALTY//SERVICE QUALITY", 6%).

The highest ranked specialty, "ACADEMIC LIBRARIES//INTERLENDING//DOCUMENT DELIVERY", focuses on library science. This category includes topics such as open access, information literacy, e-books, user needs and user behavior, interlending, library systems and reference services. The second ranked specialty, "BIBLIOMETRICS//CITATION ANALYSIS//IMPACT FACTOR", focuses on bibliometric indicators, mapping and evaluation of research and the analysis of scholarly communication. We acknowledge that a majority of the largest topics in this specialty are the same topics that were observed in the case study of *Journal of Informetrics* in the previous topics study (Sjögårde & Ahlgren, 2018, and Appendix 1 in this paper). The specialty "ENTERPRISE RESOURCE PLANNING//IT OUTSOURCING//IT INVESTMENT" includes some topics related to LIS, e.g. IT business value, IT outsourcing, Information system planning and information infrastructure. The scopes of specialties 4, 5, 6, 7, 8 and 10 are captured rather well by their labels and these specialties are all clearly related to LIS. These six specialties include information retrieval, studies of customers and service as well as library and information aspects of health service and occupation, of innovation and patents and of media and communication. The LIS relevance of "UNIVERSAL SERVICE//TELECOMMUNICATIONS//ACCESS PRICING" (rank 9) is within topics such as internet access and digital divide.

Appendix 1 lists the 10 topics with most publications in $P_{lis}$ for the top 10 ranked specialties with regard to $P_{lis}$.

Table 4: Distribution of articles in the Web of Science subject category "Information Science & Library Science" into specialties, 2011-2015.

| Rank | Specialty | # Articles in $P_{lis}$ | Total # articles in specialty | Share of specialty in $P_{lis}$ |
|---|---|---|---|---|
| 1 | ACADEMIC LIBRARIES//INTERLENDING//DOCUMENT DELIVERY | 2,779 | 3,607 | 77% |
| 2 | BIBLIOMETRICS//CITATION ANALYSIS//IMPACT FACTOR | 1,865 | 4,357 | 43% |
| 3 | ENTERPRISE RESOURCE PLANNING//IT OUTSOURCING//IT INVESTMENT | 936 | 2,512 | 37% |
| **4** | RECOMMENDER SYSTEMS//INFORMATION RETRIEVAL//COLLABORATIVE FILTERING | 756 | 6,937 | 11% |
| **5** | CUSTOMER SATISFACTION//CUSTOMER LOYALTY//SERVICE QUALITY | 562 | 9,739 | 6% |
| **6** | ELECTRONIC HEALTH RECORDS//ELECTRONIC MEDICAL RECORD//MEDICAL INFORMATICS | 487 | 3,768 | 13% |
| **7** | INNOVATION//MARKET ORIENTATION//PATENTS | 449 | 8,325 | 5% |
| **8** | KNOWLEDGE MANAGEMENT//KNOWLEDGE SHARING//OPEN SOURCE SOFTWARE | 443 | 1,537 | 29% |
| **9** | UNIVERSAL SERVICE//TELECOMMUNICATIONS//ACCESS PRICING | 331 | 1,108 | 30% |
| **10** | JOURNALISM//POLITICAL COMMUNICATION//NEWS | 291 | 5,564 | 5% |

## 6.3. The case of Medical Informatics (MI)

In analogy with the case of LIS, we retrieved all articles in $P$ that belong to a Web of Science subject category, in this case "Medical Informatics" and published in the period 2011-2015, to explore how articles within this discipline are distributed into classes in $ACPLC_S$. In total, 12,516 articles were retrieved. Let $P_{mi}$ be this set of articles.

Table 5 shows the top 10 specialties in $P_{mi}$, ranked by frequency. Only one specialty is highly concentrated into the "Medical informatics" category, namely "ELECTRONIC HEALTH RECORDS//ELECTRONIC MEDICAL RECORD//MEDICAL INFORMATICS" (which are also present in the LIS case). For the rest of the top 10 specialties, 17% or less of the total number of articles in the specialty belong to $P_{mi}$. This might suggest that MI is more interdisciplinary than LIS. It can also be the case that MI articles are published in broader journals, which are not classified into the "Medical Informatics" Web of Science subject category.

The largest specialty in "Medical Informatics" category focuses on clinical decision support systems, clinical research informatics and electronic health records. The second ranked specialty within the category, "INTERNET//MHEALTH//PERSONAL HEALTH RECORDS", addresses topics within mobile health such as personal health records, online health information and online support groups. The specialty "MISSING DATA//MULTIPLE IMPUTATION//GENERALIZED ESTIMATING EQUATIONS" includes topics related to mathematical and statistical models and methods within the medical sciences, e.g. generalized estimating equations and shared parameter models. The remainder seven top ten ranked specialties have the following foci: (4) Health technology assessment; (5) Prediction and risk models; (6) Clinical trial designs; (7) Medical epistemology, meta-analysis methods and literature searching; (8) Tele health (can be seen as a predecessor to mobile health); (9) Patient safety (includes incident and error reporting), and (10) Computational techniques such as biomedical textmining, drug target interaction and gene prioritization.

Appendix 2 lists the 10 topics with most publications in $P_{mi}$ for the top 10 ranked specialties with regard to $P_{mi}$.

Table 5: Distribution of articles in the Web of Science subject category "Medical Informatics" into specialties, 2011-2015.

| Rank | Specialty | # Articles in $P_{mi}$ | Total # articles in specialty | Share of specialty in $P_{mi}$ |
|---|---|---|---|---|
| **1** | ELECTRONIC HEALTH RECORDS//ELECTRONIC MEDICAL RECORD//MEDICAL INFORMATICS | 1544 | 3768 | 41% |
| **2** | INTERNET//MHEALTH//PERSONAL HEALTH RECORDS | 644 | 3855 | 17% |
| **3** | MISSING DATA//MULTIPLE IMPUTATION//GENERALIZED ESTIMATING EQUATIONS | 335 | 3948 | 8% |
| **4** | HEALTH TECHNOLOGY ASSESSMENT//EQ 5D//TIME TRADE OFF | 305 | 2998 | 10% |
| **5** | COMPETING RISKS//INTERVAL CENSORING//COUNTING PROCESS | 277 | 2192 | 13% |
| **6** | ADAPTIVE DESIGN//INTERIM ANALYSIS//DOSE FINDING | 267 | 1564 | 17% |
| **7** | EVIDENCE BASED MEDICINE//PUBLICATION BIAS//ABSTRACT | 203 | 3713 | 5% |
| **8** | TELEMEDICINE//TELEHEALTH//TELEPATHOLOGY | 189 | 2508 | 8% |
| **9** | PATIENT SAFETY//MEDICATION ERRORS//MEDICAL ERRORS | 187 | 3692 | 5% |
| **10** | PROTEIN INTERACTION NETWORK//PROTEIN PROTEIN INTERACTION NETWORK//GENE ONTOLOGY | 158 | 8608 | 2% |

## 7 Conclusions

In this study we have discussed how the resolution parameter given to the Modularity Optimizer software can be calibrated to cluster topics, obtained in a previous study (Sjögårde & Ahlgren, 2018) on topic identification, so that obtained publication classes correspond to the size of specialties. A set of journals has been used as a baseline for the calibration. Journals were selected based on their size and self-citation rate. The underlying assumption of our approach is that journals of a particular size and foci have a scope that correspond to specialties. By measuring the similarity between (1) the baseline classification and (2) multiple classifications obtained by using different values of the resolution parameter, we have identified a classification, which we denote as $ACPLC_s$, whose granularity corresponds to specialties.

Some criteria for the evaluation of $ACPLC_t$, the best performing ACPLC with respect to topic identification, are the same for the evaluation of $ACPLC_s$. The differences of class sizes should not be too large and “the number of very small clusters should be minimized as much as possible” (Šubelj, van Eck & Waltman, 2016). In $ACPLC_s$, 80% of the articles belong to classes consisting of 3,765-29,509 articles. Further, 80% of the articles belong to classes with a yearly publication rate of 126-1,082 articles in publication year 2006, increasing to 187-2,040 in the publication year 2015. Only 0.9% of the articles in $P$ belong to classes with a total number of articles of less than 500. As in the previous study, the distribution follows a typical scientometric distribution, and we therefore consider the results, regarding class sizes, as satisfying.

In accordance with the previous study, we stress that it is reasonable of practical reasons to reassign small classes, which we have not done in this study. Moreover, we consider content labelling of classes as a topic that needs to be addressed in future work.

Another criteria stated by Šubelj, van Eck & Waltman (2016) is that classes should make intuitive sense. In addition, we stress that the focus of a specialty should be possible to identify and that two specialties should have subject foci that can be distinguished. Two case studies, in which we have identified specialties within the disciplines LIS and MI, have been performed to evaluate these criteria. We could identify the subject foci of the specialties in these case studies, and the subject foci of the specialties have been relatively easy to distinguish. Thus, the two criteria are (approximately) satisfied in our case. Further, several of the specialties identified in the LIS case study have been identified by others (Bauer, Leydesdorff, & Bornmann, 2016; Blessinger & Frasier, n.d.; Figuerola, García Marco, & Pinto, 2017; Janssens, Leta, Glänzel, & De Moor, 2006) and the same holds for several of the specialties identified in the MI case study (Kim & Delen, 2018; Schuemie, Talmon, Moorman, & Kors, 2009; Wang, Topaz, Plasek, & Zhou, 2017). However, more case studies are needed to verify the soundness of the used methodology.

The beforementioned feature of the classification approach used in this study, logical classification, which assigns each topic to exactly one speciality, has some limitations. It is clear that topics can be addressed by several specialties (or at a higher level disciplines). For instance, Appendix 1 and 2 show that the topic with the label "NATURAL LANGUAGE PROCESSING//MEDICAL LANGUAGE PROCESSING//CLINICAL TEXT" is addressed by both the LIS and the MI discipline. This topic is forced into exactly one specialty, "ELECTRONIC HEALTH RECORDS//ELECTRONIC MEDICAL RECORD//MEDICAL INFORMATICS". Thus, relations between this topic and e.g. specialties within the LIS discipline is not expressed by $ACPLC_s$. However, relations between a specialty and topics within other specialties can still be analyzed using, for instance, citation relations. Nevertheless, a logical classification to some extent oversimplifies the complex structure of topic representation in research publications.

The classification of a topic into a specialty may also be counter-intuitive from the point of view of a single researcher active in one of the involved specialties or disciplines. An advantage of logical classification is, however, that such a classification might put topics and specialties in a larger context. As an example, the technology acceptance model (TAM) is an information system theory used within LIS. However, TAM is not only used within LIS, but also by other fields, e.g. computer science. In $ACPLC_s$, the topic "TECHNOLOGY ACCEPTANCE MODEL//TECHNOLOGY ACCEPTANCE MODEL TAM//TAM" is categorized in the specialty "CUSTOMER SATISFACTION//CUSTOMER LOYALTY//SERVICE QUALITY", and thereby $ACPLC_s$ puts LIS publications within this topic in a larger context.

The combined outcome of our previous study on the classification of topics, and the present study on the classification of specialties, is a two-level hierarchical classification. We believe that such classification comprises a valuable part of a research information system and propose that such classification can be used for bibliometric analyses of topics and specialties.

# Appendix 1: Topics per specialty – LIS

| SPECIALTY<br>TOPIC | # Articles in $P_{lis}$ | # Articles in topic | Shr. of topic in $P_{lis}$ |
|---|---|---|---|
| **ACADEMIC LIBRARIES//INTERLENDING//DOCUMENT DELIVERY** | | | |
| INFORMATION LITERACY//INFORMATION LITERACY INSTRUCTION//LIBRARY INSTRUCTION | 194 | 212 | 92% |
| ELECTRONIC BOOKS//E BOOKS//E TEXTBOOK | 134 | 172 | 78% |
| OPEN ACCESS//OPEN ACCESS JOURNALS//GOLD OPEN ACCESS | 123 | 218 | 56% |
| DOCUMENT DELIVERY//INTERLENDING//INTERLIBRARY LOAN | 121 | 122 | 99% |
| LIBRARY 20//LIBRARIAN 2//ACADEMIC LIBRARIES | 104 | 110 | 95% |
| KNOWLEDGE ORGANIZATION//FACETED CLASSIFICATIONS//INDEXING LANGUAGE | 93 | 106 | 88% |
| INTERACTIVE INFORMATION RETRIEVAL//END USER SEARCHING//INFORMATION NEEDS AND USES | 92 | 128 | 72% |
| INFORMATION PRACTICES//AIDS TALK//BARRIERS TO INFORMATION SEEKING | 85 | 99 | 86% |
| INFORMATION SCIENCE//DIKW HIERARCHY//PROPERTIES OF DOCUMENTARY PRACTICE | 81 | 99 | 82% |
| ELECTRONIC JOURNALS//ELECTRONIC PERIODICALS//E JOURNALS | 79 | 85 | 93% |
| **BIBLIOMETRICS//CITATION ANALYSIS//IMPACT FACTOR** | | | |
| FIELD NORMALIZATION//SOURCE NORMALIZATION//RESEARCH EVALUATION | 193 | 258 | 75% |
| H INDEX//HIRSCH INDEX//G INDEX | 190 | 317 | 60% |
| RESEARCH COLLABORATION//SCIENTIFIC COLLABORATION//CO AUTHORSHIP | 125 | 184 | 68% |
| AUTHOR CO CITATION ANALYSIS//BIBLIOGRAPHIC COUPLING//CO CITATION ANALYSIS | 83 | 142 | 58% |
| GOOGLE SCHOLAR//SCOPUS//WEB OF SCIENCE | 79 | 121 | 65% |
| ALTMETRICS//MENDELEY//RESEARCHGATE | 75 | 113 | 66% |
| OVERLAY MAP//SCIENCE OVERLAY MAPS//JOURNAL CLASSIFICATION | 75 | 148 | 51% |
| CO AUTHORSHIP NETWORKS//SCIENTIFIC COLLABORATION//CO AUTHOR NETWORKS | 70 | 136 | 51% |
| BOOK CITATION INDEX//SOCIAL SCIENCES AND HUMANITIES//BOOK PUBLISHERS | 65 | 93 | 70% |
| WEBOMETRICS//WEB VISIBILITY//WEB LINKS | 60 | 80 | 75% |
| **ENTERPRISE RESOURCE PLANNING//IT OUTSOURCING//IT INVESTMENT** | | | |
| IT BUSINESS VALUE//BUSINESS VALUE OF IT//IT INVESTMENT | 92 | 184 | 50% |
| STRATEGIC INFORMATION SYSTEMS PLANNING//CHIEF INFORMATION OFFICER//IT GOVERNANCE | 70 | 124 | 56% |
| IT OUTSOURCING//OUTSOURCING//IS OUTSOURCING | 70 | 159 | 44% |
| IS RESEARCH//REFERENCE DISCIPLINE//IS DISCIPLINE | 65 | 83 | 78% |
| TECHNOLOGICAL FRAMES//STRUCTURATION THEORY//USER RESISTANCE | 57 | 102 | 56% |
| SOCIOMATERIALITY//DIGITAL INNOVATION//LAYERED MODULAR ARCHITECTURE | 45 | 71 | 63% |
| CRITICAL RESEARCH//IS FIELD//INTERPRETIVE RESEARCH | 41 | 59 | 69% |
| ENTERPRISE RESOURCE PLANNING//ENTERPRISE RESOURCE PLANNING ERP//ERP IMPLEMENTATION | 41 | 196 | 21% |
| TOE FRAMEWORK//E COMMERCE ADOPTION//TECHNOLOGY ORGANIZATION ENVIRONMENT FRAMEWORK | 38 | 136 | 28% |
| DESIGN SCIENCE//DESIGN SCIENCE RESEARCH//DESIGN THEORIZING | 34 | 81 | 42% |
| **RECOMMENDER SYSTEMS//INFORMATION RETRIEVAL//COLLABORATIVE FILTERING** | | | |
| FOLKSONOMY//SOCIAL TAGGING//COLLABORATIVE TAGGING | 57 | 159 | 36% |
| SENTIMENT ANALYSIS//OPINION MINING//SENTIMENT CLASSIFICATION | 35 | 371 | 9% |
| AUTHOR NAME DISAMBIGUATION//AUTHOR DISAMBIGUATION//NAME DISAMBIGUATION | 34 | 66 | 52% |
| RECOMMENDER SYSTEMS//COLLABORATIVE FILTERING//RECOMMENDATION SYSTEM | 30 | 576 | 5% |
| RELEVANCE CRITERIA//RELEVANCE JUDGEMENT//TEST COLLECTIONS | 25 | 49 | 51% |
| SESSION IDENTIFICATION//QUERY LOG ANALYSIS//QUERY RECOMMENDATION | 24 | 86 | 28% |
| COMMUNITY QUESTION ANSWERING//SOCIAL QA//ANSWER RECOMMENDATION | 23 | 54 | 43% |
| COLLABORATIVE INFORMATION SEEKING//SEARCH HISTORIES//SOCIAL SEARCH | 22 | 40 | 55% |
| EXPERT FINDING//EXPERT SEARCH//ENTITY RETRIEVAL | 20 | 87 | 23% |
| MULTI DOCUMENT SUMMARIZATION//TEXT SUMMARIZATION//DOCUMENT SUMMARIZATION | 16 | 146 | 11% |
| **CUSTOMER SATISFACTION//CUSTOMER LOYALTY//SERVICE QUALITY** | | | |
| TECHNOLOGY ACCEPTANCE MODEL//TECHNOLOGY ACCEPTANCE MODEL TAM//TAM | 88 | 551 | 16% |
| ONLINE REVIEWS//ELECTRONIC WORD OF MOUTH//WORD OF MOUTH | 53 | 451 | 12% |
| CONTINUANCE INTENTION//IS CONTINUANCE//EXPECTATION CONFIRMATION MODEL | 47 | 158 | 30% |
| TRUST TRANSFER//ONLINE TRUST//INITIAL TRUST | 40 | 164 | 24% |
| SOCIAL MEDIA//ACTIONABLE INTELLIGENCE//BEST TIME TO TWEET | 35 | 137 | 26% |
| BRAND COMMUNITY//ONLINE BRAND COMMUNITY//VIRTUAL COMMUNITIES | 33 | 173 | 19% |
| FORMATIVE MEASUREMENT//CAUSAL INDICATORS//FORMATIVE INDICATORS | 31 | 125 | 25% |
| E SATISFACTION//E SERVICE QUALITY//WEBSITE QUALITY | 26 | 154 | 17% |
| RECOMMENDATION AGENT//RECOMMENDATION AGENTS//INTERACTIVE DECISION AIDS | 16 | 50 | 32% |
| SPONSORED SEARCH//GENERALIZED SECOND PRICE AUCTION//SPONSORED SEARCH AUCTIONS | 12 | 193 | 6% |
| **ELECTRONIC HEALTH RECORDS//ELECTRONIC MEDICAL RECORD//MEDICAL INFORMATICS** | | | |
| NATURAL LANGUAGE PROCESSING//MEDICAL LANGUAGE PROCESSING//CLINICAL TEXT | 92 | 278 | 33% |
| HEALTH INFORMATION TECHNOLOGY//ELECTRONIC HEALTH RECORDS//MEANINGFUL USE | 46 | 392 | 12% |
| ALERT FATIGUE//CLINICAL DECISION SUPPORT SYSTEMS//CLINICAL DECISION SUPPORT | 43 | 216 | 20% |
| CDISC//ISO IEC 11179//CLINICAL RESEARCH INFORMATICS | 37 | 167 | 22% |
| PHEWAS//PHENOME WIDE ASSOCIATION STUDY//CLINICAL PHENOTYPE MODELING | 34 | 160 | 21% |
| HEALTH INFORMATION EXCHANGE//HEALTH RECORD BANK//REGIONAL HEALTH INFORMATION ORGANIZATIONS | 32 | 154 | 21% |
| CPOE//E PRESCRIBING//ELECTRONIC PRESCRIBING | 31 | 218 | 14% |
| OPENEHR//LOINC//CLINICAL ARCHETYPES | 21 | 111 | 19% |
| SNOMED CT//UMLS//ABSTRACTION NETWORK | 13 | 108 | 12% |
| COPY PASTE//CLINICAL DOCUMENTATION//COMPUTER BASED DOCUMENTATION | 10 | 59 | 17% |
| **INNOVATION//MARKET ORIENTATION//PATENTS** | | | |
| PATENT ANALYSIS//PATENT MINING//TECHNOLOGY INTELLIGENCE | 41 | 179 | 23% |
| NON PATENT REFERENCES//NON PATENT CITATION//SCIENCE LINKAGE | 34 | 57 | 60% |
| PROBABILISTIC ENTROPY//UNIVERSITY INDUSTRY GOVERNMENT RELATIONSHIP//TRIPLE HELIX | 32 | 52 | 62% |
| ACADEMIC ENTREPRENEURSHIP//ENTREPRENEURIAL UNIVERSITY//UNIVERSITY SPIN OFFS | 27 | 421 | 6% |
| PATENT VALUE//PATENTS//PATENT SYSTEM | 27 | 202 | 13% |
| USER INNOVATION//LEAD USERS//INNOVATION CONTESTS | 19 | 226 | 8% |
| SOFTWARE ECOSYSTEMS//BUSINESS ECOSYSTEM//MOBILE COMPUTING INDUSTRY | 16 | 81 | 20% |
| STRUCTURAL HOLES//ADVICE RELATIONS//BROKERAGE | 15 | 178 | 8% |
| AMBIDEXTERITY//ORGANIZATIONAL AMBIDEXTERITY//EXPLOITATION | 14 | 232 | 6% |
| ABSORPTIVE CAPACITY//POTENTIAL ABSORPTIVE CAPACITY//COMBINATIVE CAPABILITIES | 13 | 102 | 13% |
| **KNOWLEDGE MANAGEMENT//KNOWLEDGE SHARING//OPEN SOURCE SOFTWARE** | | | |
| KNOWLEDGE SHARING//KNOWLEDGE MANAGEMENT//KNOWLEDGE SHARING BEHAVIOR | 151 | 328 | 46% |
| KNOWLEDGE MANAGEMENT//ENTERPRISE BENEFITS//KNOWLEDGE CHAIN | 58 | 101 | 57% |
| OPEN SOURCE SOFTWARE//OPEN SOURCE//OPEN SOURCE SOFTWARE OSS | 53 | 230 | 23% |
| WIKIPEDIA//COOPERATIVE KNOWLEDGE GENERATION//ENCYCLOPAEDIAS | 22 | 76 | 29% |
| COMMUNITIES OF PRACTICE//ORGANIZING PRACTICES//COMMUNITY OF PRACTICE | 19 | 82 | 23% |
| INTELLECTUAL CAPITAL//INTANGIBLE ASSETS//INTELLECTUAL CAPITAL IC | 19 | 96 | 20% |
| ENTERPRISE EVOLUTION//KNOWLEDGE CREATION//AUTOMOBILE PROJECT | 16 | 43 | 37% |
| EUROPEAN SMES//BARRIERS OF IMPLEMENTATION//CASE STUDY IN SINGAPORE | 13 | 35 | 37% |
| BLACK HAT SEO//COMMUNICATIONS ACTIVITIES//CONSUMER COMPARISON | 11 | 28 | 39% |
| INFORMATION CULTURE//AUXILIARY ROCKETS//GASEOUS PROPELLANT ROCKET MOTORS | 7 | 14 | 50% |
| **UNIVERSAL SERVICE//TELECOMMUNICATIONS//ACCESS PRICING** | | | |
| DIGITAL DIVIDE//BROADBAND ADOPTION//BROADBAND | 71 | 150 | 47% |
| ACCESS REGULATION//ACCESS PRICING//NEXT GENERATION ACCESS NETWORKS | 48 | 121 | 40% |
| TD SCDMA//FORMAL STANDARDS//WAPI | 39 | 63 | 62% |
| SPECTRUM AUCTIONS//DIGITAL DIVIDEND//SPECTRUM TRADING | 28 | 68 | 41% |
| FIXED MOBILE SUBSTITUTION//MOBILE TELECOMMUNICATIONS//FIXED TO MOBILE SUBSTITUTION | 24 | 56 | 43% |
| BILL AND KEEP//TERMINATION RATES//ACCESS PRICING | 17 | 60 | 28% |
| UNIVERSAL SERVICE//E RATE//UNIVERSAL SERVICE FUND | 17 | 35 | 49% |
| NET NEUTRALITY//NETWORK NEUTRALITY//CONTENT PROVIDERS | 17 | 70 | 24% |
| PRICE CAPS//INCENTIVE REGULATION//PRICE CAP REGULATION | 12 | 39 | 31% |
| MARKET INFORMATION SERVICES//AFRICA AS A SPECIAL CASE//BAROQUIZATION | 10 | 91 | 11% |
| **JOURNALISM//POLITICAL COMMUNICATION//NEWS** | | | |
| MEDIA CREDIBILITY//MEDIA SKEPTICISM//CREDIBILITY | 41 | 104 | 39% |
| POLITICAL PARTICIPATION//ONLINE POLITICAL PARTICIPATION//SOCIAL MEDIA | 32 | 298 | 11% |
| USES AND GRATIFICATIONS//ONLINE VIDEO SERVICES//GRATIFICATIONS | 22 | 94 | 23% |
| TWITTER//HASHTAGS//ELECTION DEBATE | 19 | 125 | 15% |
| ONLINE JOURNALISM//CITIZEN JOURNALISM//JOURNALISM | 17 | 198 | 9% |
| ONLINE CAMPAIGNING//PARTY WEBSITES//INTERNET POLITICS | 15 | 112 | 13% |
| HYPERLINK NETWORK//HYPERLINK ANALYSIS//INTERNATIONAL HYPERLINK | 12 | 57 | 21% |
| SELECTIVE EXPOSURE//ONE STEP FLOW//PARTISAN MEDIA | 12 | 117 | 10% |
| COMMENTS IN NEWS//ONLINE COMMENTS//ONLINE DELIBERATION | 10 | 84 | 12% |
| SENSATIONALISM//INTERPRETIVE COMPLEXITY//ANTIDRUG PUBLIC SERVICE ANNOUNCEMENTS | 8 | 78 | 10% |

## Appendix 2: Topics per specialty – MI

| SPECIALTY<br>TOPIC | # Articles in $P_{mi}$ | # Articles in topic | Shr. of topic in $P_{mi}$ |
|---|---|---|---|
| **ELECTRONIC HEALTH RECORDS//ELECTRONIC MEDICAL RECORD//MEDICAL INFORMATICS** | | | |
| NATURAL LANGUAGE PROCESSING//MEDICAL LANGUAGE PROCESSING//CLINICAL TEXT | 188 | 278 | 68% |
| HEALTH INFORMATION TECHNOLOGY//ELECTRONIC HEALTH RECORDS//MEANINGFUL USE | 127 | 392 | 32% |
| CDISC//ISO IEC 11179//CLINICAL RESEARCH INFORMATICS | 112 | 167 | 67% |
| ALERT FATIGUE//CLINICAL DECISION SUPPORT SYSTEMS//CLINICAL DECISION SUPPORT | 110 | 216 | 51% |
| NURSING INFORMATION SYSTEM//CLINICAL INFORMATION SYSTEMS//DOCUMENTATION TIME | 104 | 152 | 68% |
| OPENEHR//LOINC//CLINICAL ARCHETYPES | 89 | 111 | 80% |
| HEALTH INFORMATION EXCHANGE//HEALTH RECORD BANK//REGIONAL HEALTH INFORMATION ORGANIZATIONS | 87 | 154 | 56% |
| CPOE//E PRESCRIBING//ELECTRONIC PRESCRIBING | 84 | 218 | 39% |
| SNOMED CT//UMLS//ABSTRACTION NETWORK | 76 | 108 | 70% |
| PHEWAS//PHENOME WIDE ASSOCIATION STUDY//CLINICAL PHENOTYPE MODELING | 49 | 160 | 31% |
| **INTERNET//MHEALTH//PERSONAL HEALTH RECORDS** | | | |
| PERSONAL HEALTH RECORDS//PATIENT PORTAL//PATIENT ACCESS TO RECORDS | 112 | 281 | 40% |
| INTERNET//HEALTH INFORMATION//ONLINE HEALTH INFORMATION | 48 | 193 | 25% |
| ONLINE SUPPORT GROUPS//COMPREHENSIVE HEALTH ENHANCEMENT SUPPORT SYSTEM CHESS//INTERNET CANCER SUPPORT GROUPS | 37 | 220 | 17% |
| MEDICAL APP//APPS//SMARTPHONE | 33 | 165 | 20% |
| MHEALTH//MEDICATION REMINDERS//REAL TIME ADHERENCE MONITORING | 32 | 275 | 12% |
| TWITTER MESSAGING//INFOVEILLANCE//INFODEMIOLOGY | 31 | 113 | 27% |
| E PROFESSIONALISM//SOCIAL MEDIA//TWITTER MESSAGING | 30 | 274 | 11% |
| TEXT MESSAGING//TEXT MESSAGE//MHEALTH | 29 | 218 | 13% |
| MOBILE APPS//APPS//MHEALTH | 25 | 126 | 20% |
| PHYSICIAN RATING WEBSITE//RATING SITES//QUALITY TRANSPARENCY | 23 | 61 | 38% |
| **MISSING DATA//MULTIPLE IMPUTATION//GENERALIZED ESTIMATING EQUATIONS** | | | |
| GENERALIZED ESTIMATING EQUATIONS//QUASI LEAST SQUARES//GEE | 23 | 104 | 22% |
| MULTIPLE IMPUTATION//MISSING DATA//PREDICTIVE MEAN MATCHING | 23 | 218 | 11% |
| JOINT MODEL//SHARED PARAMETER MODEL//DYNAMIC PREDICTIONS | 22 | 112 | 20% |
| PATTERN MIXTURE MODEL//MISSING NOT AT RANDOM//MISSING DATA | 19 | 137 | 14% |
| CLUSTER RANDOMIZED TRIAL//CLUSTER RANDOMIZATION//GROUP RANDOMIZED TRIAL | 16 | 126 | 13% |
| ZERO INFLATION//ZERO INFLATED MODELS//OVERDISPERSION | 16 | 142 | 11% |
| CENSORED COVARIATE//CENSORED PREDICTOR//TWO PART STATISTICS | 13 | 41 | 32% |
| REGRESSION CALIBRATION//MEASUREMENT ERROR//CORRECTED SCORE | 12 | 105 | 11% |
| MONOTONE MISSING//DISCRETE TIME LONGITUDINAL DATA//INDEPENDENT MISSING | 12 | 37 | 32% |
| DOUBLE ROBUSTNESS//AUGMENTED INVERSE PROBABILITY WEIGHTING AIPW//MISSING AT RANDOM | 10 | 67 | 15% |
| **HEALTH TECHNOLOGY ASSESSMENT//EQ 5D//TIME TRADE OFF** | | | |
| HEALTH TECHNOLOGY ASSESSMENT//HOSPITAL BASED HTA//MINI HTA | 58 | 118 | 49% |
| EQ 5D//SF 6D//EQ 5D 5L | 38 | 411 | 9% |
| VALUE OF INFORMATION//OPTIMAL TRIAL DESIGN//VALUE OF INFORMATION ANALYSIS | 29 | 95 | 31% |
| HEALTH TECHNOLOGY ASSESSMENT//INSTITUTE FOR QUALITY AND EFFICIENCY IN HEALTH CARE//FOURTH HURDLE | 28 | 153 | 18% |
| DYNAMIC TRANSMISSION//HALF CYCLE CORRECTION//COST EFFECTIVENESS MODELING | 26 | 82 | 32% |
| STRENGTH OF PREFERENCES//IN PERSON INTERVIEW//MULTI CRITERIA DECISION ANALYSIS | 15 | 67 | 22% |
| COST EFFECTIVENESS RATIOS//NET HEALTH BENEFIT//COST EFFECTIVENESS ACCEPTABILITY CURVES | 14 | 74 | 19% |
| COVERAGE WITH EVIDENCE DEVELOPMENT//MEDICARE COVERAGE//RISK SHARING AGREEMENTS | 14 | 91 | 15% |
| HORIZON SCANNING SYSTEMS//HORIZON SCANNING//EARLY AWARENESS AND ALERT SYSTEMS | 12 | 20 | 60% |
| STANDARD GAMBLE//TIME TRADE OFF//UTILITY ASSESSMENT | 9 | 81 | 11% |
| **COMPETING RISKS//INTERVAL CENSORING//COUNTING PROCESS** | | | |
| INTEGRATED DISCRIMINATION IMPROVEMENT//NET RECLASSIFICATION IMPROVEMENT//DECISION ANALYTIC MEASURES | 28 | 110 | 25% |
| MULTISTATE MODEL//ILLNESS DEATH PROCESS//AALEN JOHANSEN ESTIMATOR | 23 | 111 | 21% |
| COMPETING RISKS//CUMULATIVE INCIDENCE FUNCTION//CAUSE SPECIFIC HAZARD | 20 | 117 | 17% |
| RECURRENT EVENTS//PANEL COUNT DATA//INFORMATIVE OBSERVATION TIMES | 19 | 174 | 11% |
| EXPLAINED VARIATION//TIME DEPENDENT ROC//C INDEX | 19 | 68 | 28% |
| CURE RATE MODEL//CURE MODEL//LONG TERM SURVIVAL MODELS | 17 | 95 | 18% |
| SURROGATE ENDPOINT//PRENTICE CRITERION//LIKELIHOOD REDUCTION FACTOR | 13 | 84 | 15% |
| INTERVAL CENSORING//CURRENT STATUS DATA//INTERVAL CENSORED DATA | 13 | 127 | 10% |
| CASE COHORT DESIGN//CASE COHORT//CASE COHORT STUDY | 12 | 77 | 16% |
| FRAILTY MODEL//CORRELATED FAILURE TIMES//CROSS RATIO FUNCTION | 12 | 104 | 12% |
| **ADAPTIVE DESIGN//INTERIM ANALYSIS//DOSE FINDING** | | | |
| ADAPTIVE DESIGN//GROUP SEQUENTIAL TEST//GROUP SEQUENTIAL DESIGN | 58 | 221 | 26% |
| CONTINUAL REASSESSMENT METHOD//DOSE FINDING//DOSE FINDING STUDIES | 50 | 200 | 25% |
| TWO STAGE DESIGN//PHASE II DESIGN//PHASE II CLINICAL TRIALS | 28 | 113 | 25% |
| FAMILYWISE ERROR RATE//GATEKEEPING PROCEDURE//MULTIPLE TESTS | 27 | 119 | 23% |
| SCORE INTERVAL//BINOMIAL PROPORTION//BINOMIAL DISTRIBUTION | 18 | 140 | 13% |
| NONINFERIORITY MARGIN//NON INFERIORITY//NON INFERIORITY TRIAL | 16 | 112 | 14% |
| MINIMUM EFFECTIVE DOSE//MCP MOD//WILLIAMS TEST | 12 | 57 | 21% |
| META ANALYTIC PREDICTIVE//EPSILON INFORMATION PRIOR//COMPUTATIONALLY INTENSIVE METHODS | 9 | 42 | 21% |
| MULTIREGIONAL CLINICAL TRIAL//BRIDGING STUDY//MULTIREGIONAL TRIAL | 8 | 68 | 12% |
| AVERAGE COVERAGE CRITERION//AVERAGE LENGTH CRITERION//BAYESIAN POINT OF VIEW | 7 | 57 | 12% |
| **EVIDENCE BASED MEDICINE//PUBLICATION BIAS//ABSTRACT** | | | |
| MULTIVARIATE META ANALYSIS//DERSIMONIAN LAIRD ESTIMATOR//MANDEL PAULE ALGORITHM | 40 | 150 | 27% |
| MEDICAL EPISTEMOLOGY//EVIDENCE BASED MEDICINE//EVIDENCE IN MEDICINE | 32 | 77 | 42% |
| MIXED TREATMENT COMPARISON//NETWORK META ANALYSIS//MULTIPLE TREATMENTS META ANALYSIS | 26 | 198 | 13% |
| MEDLINE//EMBASE//LITERATURE SEARCHING | 11 | 112 | 10% |
| NUMBER NEEDED TO TREAT//ABSOLUTE RISK REDUCTION//NUMBER NEEDED TO TREAT NNT | 9 | 52 | 17% |
| TRIAL REGISTRATION//CLINICALTRIALSGOV//PUBLICATION BIAS | 7 | 250 | 3% |
| PUBLICATION BIAS//FUNNEL PLOT//SMALL STUDY EFFECTS | 6 | 64 | 9% |
| JOURNAL CLUB//EVIDENCE BASED MEDICINE EDUCATION//FRESNO TEST | 6 | 155 | 4% |
| AWARENESS SCORE//CHIROPRACTIC QUESTIONNAIRES//COMMUNITY OF PRACTICE KNOWLEDGE NETWORKS | 6 | 45 | 13% |
| CONFLICT OF INTEREST//EDITORIAL ETHICS//CONFLICTS OF INTEREST | 6 | 189 | 3% |
| **TELEMEDICINE//TELEHEALTH//TELEPATHOLOGY** | | | |
| TELEHEALTH//TELECARE//TELEHEALTHCARE | 32 | 188 | 17% |
| TELEMONITORING//HOME TELEMONITORING//TETEMONITORING | 27 | 155 | 17% |
| ELDERCARE TECHNOLOGY//HOME BASED CLINICAL ASSESSMENT//PASSIVE INFRARED PIR MOTION DETECTORS | 19 | 90 | 21% |
| TELE ECHOGRAPHY//TELESONOGRAPHY//TELE ULTRASOUND | 10 | 48 | 21% |
| MOBILE TELEMEDICINE//MOBILE CARE//TIME FREQUENCY ENERGY DISTRIBUTIONS | 7 | 34 | 21% |
| CHRONIC DISEASE METHODS THERAPY//CRITICAL PATHWAYS MESH//HEATH CARE PRACTICES | 7 | 24 | 29% |
| TELEREHABILITATION//TELEPRACTICE//REMOTE ASSESSMENT | 6 | 90 | 7% |
| TELE EEG//INITIATE BUILD OPERATE TRANSFER STRATEGY//INTERNATIONAL VIRTUAL E HOSPITAL FOUNDATION | 5 | 69 | 7% |
| ISI WEB OF SCIENCE DATABASE//LISTENING STYLES//NON ADHERENCE FACTORS | 5 | 24 | 21% |
| TELEDERMATOLOGY//MOBILE TELEDERMATOLOGY//STORE AND FORWARD | 4 | 135 | 3% |
| **PATIENT SAFETY//MEDICATION ERRORS//MEDICAL ERRORS** | | | |
| MEDICATION ERRORS//SMART PUMPS//MEDICATION ADMINISTRATION ERRORS | 25 | 281 | 9% |
| BAR CODE MEDICATION ADMINISTRATION//BAR CODED MEDICATION ADMINISTRATION//SCANNING COMPLIANCE | 24 | 82 | 29% |
| SIGN OUT//HANDOFF//HANDOVER | 19 | 334 | 6% |
| VOCERA//HOSPITAL COMMUNICATION SYSTEMS//PAGERS | 19 | 66 | 29% |
| INTERRUPTION//DISTRACTIONS//TASK SEVERITY | 13 | 162 | 8% |
| MEDWISE//THREAT AND ERROR MANAGEMENT TEM//USER CONFIGURABLE EHR | 11 | 32 | 34% |
| INCIDENT REPORTING//MEDICATION INCIDENTS//ERROR REPORTING | 8 | 155 | 5% |
| MEDICAL DEVICE DESIGN//INSTITUTIONAL DECISION MAKING//USER COMPUTER | 8 | 34 | 24% |
| NON TECHNICAL SKILLS//TEAMWORK//TEAM TRAINING | 7 | 317 | 2% |
| TRIGGER TOOL//GLOBAL TRIGGER TOOL//PREVENTABLE HARM | 6 | 182 | 3% |
| **PROTEIN INTERACTION NETWORK//PROTEIN PROTEIN INTERACTION NETWORK//GENE ONTOLOGY** | | | |
| BIOMEDICAL TEXT MINING//TEXT MINING//BIOMEDICAL NAMED ENTITY RECOGNITION | 46 | 368 | 13% |
| DRUG TARGET INTERACTION//DRUG REPOSITIONING//DRUG TARGET INTERACTION PREDICTION | 14 | 353 | 4% |
| LITERATURE BASED DISCOVERY//LITERATURE RELATED DISCOVERY//SEMANTIC FILTERS | 12 | 65 | 18% |
| CONTEXT CONSTRAINED NETWORKS//FUNCTIONAL NODES//MASTER INTEGRATOR | 8 | 17 | 47% |
| GENE PRIORITIZATION//HUMAN DISEASE NETWORK//DISEASE GENE PREDICTION | 8 | 225 | 4% |
| MATCHMAKER EXCHANGE//COMPARATIVE PHENOMICS//GENOMIC API | 7 | 222 | 3% |
| PATHOGENIC GENE SELECTION//APPLICATION IN BIOMEDICINE//AUTOMATIC MASS ASSIGNMENT | 6 | 123 | 5% |
| GENE ONTOLOGY//SEMANTIC SIMILARITY//SEMANTIC SIMILARITY MEASURE | 5 | 130 | 4% |
| SPARQL//FAIR DATA//SEMANTIC WEB | 4 | 89 | 4% |
| GENE SET ANALYSIS//GENE SET ENRICHMENT//GENE SET TESTING | 4 | 209 | 2% |